\documentclass{article}

\usepackage{arxiv}

\usepackage[utf8]{inputenc} 
\usepackage[T1]{fontenc}    
\usepackage{hyperref}       
\usepackage{url}            
\usepackage{booktabs}       
\usepackage{amsfonts}       
\usepackage{nicefrac}       
\usepackage{microtype}      
\usepackage{lipsum}		
\usepackage{graphicx}
\usepackage{natbib}
\usepackage{doi}
\usepackage{mathtools}
\usepackage{multirow}
\usepackage{adjustbox}
\newcommand{\e}{\begin{equation}}
\newcommand{\ee}{\end{equation}}
\def\b{\ensuremath\boldsymbol}
\def\mc{\ensuremath\mathcal}

\newtheorem{theorem}{Theorem}

\newtheorem{remark}{Remark}

\title{Unique Distributions Under Non-IID Assumption}

\author{ \href{https://orcid.org/0000-0002-3289-8922}{\hspace{1mm}Kali P.~Chowdhury}\thanks{I would especially like to acknowledge Laura Smith, Weining Shen and Knut Solna for their invaluable insights into the various theorems and their applications. I would also like to thank the University of California, Irvine's Machine Learning Lab for their generous access to the data.} \\
Johns Hopkins University\\
	\texttt{dukechowdhury@icloud.com} \\
}

\date{}

\renewcommand{\shorttitle}{\textit{arXiv}}

\hypersetup{
	pdftitle={Nonparametric Functional Analysis of Generalized Linear Models Under Nonlinear Constraints},
	pdfsubject={},
	pdfauthor={K. P. Chowdhury},
	pdfkeywords={Unique Distributions; Unbalanced Data; MCMC; Artificial Intelligence; Machine Learning; Nonparametric Regression and Categorical Data Analysis, Model Diagnostics; Model Fit, Inference and Prediction.},
}

\begin{document}
	\maketitle
	
	\begin{abstract}
			Applications of Strongly Convergent M-Estimators are discussed. Given the ubiquity of distributions across the sciences, multiple applications in the Physical, Biomedical and Social Sciences are elaborated. In one particular implementation unique utilities are attained. Finally, the importance of the results and findings to model fit, inference and prediction are highlighted for broad applicability across the sciences. 
\end{abstract}

\section{Introduction}
In \citep{chow21, nonparafuncarxiv, stronglyconvergentMestarxiv} various mathematical foundations are given for unique M-estimators under very minimal assumptions, without any need for either independent or identically distributed observations. This has ready extensions to numerous applications across the sciences, including in ordered and unordered models in either univariate or multivariate formulation. Since the theoretical results have already been presented, in this paper I briefly discuss simulation results and some realworld data application outcomes.

Thus, consider any mathematical model where we seek to specify a relationship between between two variables of interest such as, \e\b y_i = c(\b x_i)'\b \beta, \ee 
where the $ c(X_i) $ are some desired continuous functions of the observed design variables (covariates). We seek to estimate a functional specification for $ y_i $ in regards to the apriori functional specification of $ c(\b x_i) $. We will show that such an apriori specification allows us to nonparametrically find the true unique distribution of $ \b y $ without the need for any restrictive assumptions on our model specification. A deeper introspection will further show that an apriori model specification is in fact a strength of this methodology, since we may make the apriori specifications more complex as needed. They key being that as a function of this apriori specification we can compute the usual scientific parameters of interest $\b \beta $ (which may also be a matrix).

Since the cornerstone of statistical models are the various forms of independent and identicaly distributed (IID) random variables, in this paper I consider methods when such IID assumptions need not hold. I further discuss why the existing framework in the absence of such an IID framework make asymptotic inferences impractical. Thus, to that end I further ensure that this construction gives results identical to existing constructs should traditional assumptions hold. Of particular interest is the convergence specifications under which scientific inferences are made for such models under varying assumptions on the underlying model, and how this affects estimation of strongly convergent unbiased parameters of interst for stochastic models.

As such, consider the latent variable and non-latent variable frameworks, where in the existing latent variable formulations pointwise convergence is not guaranteed for the methods even under very strong parametric assumptions on the probability of success across such methodologies (\cite{chow21}, \cite{sdss21}). Accordingly, let the probability of successes be $ F $ and $ F^* $ for binary regression and latent variable formulations respectively, which are necessarily unknown. The forthcoming discussion will make clear that the link constraint holding for each observation in this framework is an absolutely crucial component needed for almost sure convergence to hold. This assertion is true no matter the estimation technique involved such as \cite{tanner1987calculation} or MLE since the characteristics of the underlying imposed topology are crucial for a properly defined linear operator on the relevant abelian group. Thus, in the forthcoming Section \ref{mathfoundations}, I present the theoretical foundations referring the reader to the relevant works for more detailed proofs. In Section \ref{simulations} I present extensive simulation results for the methodology. In Section \ref{application} various applications to the Biomedical, Physical Science and the Social Sciences are elaborated. These applications are discussed further in Section \ref{unifdisc} and finally I present the concluding thoughts in Section \ref{conclusion}.

\section{Methodology}\label{mathfoundations}

\cite{stronglyconvergentMestarxiv} presents an unifying framework under which robust M-estimators may be constructed which converges almost surely to the true parameters. Here I briefly highlight the methodology before going on to the empirical applications.

	%
	\subsection{An Unified Almost Sure Convergence Methodology}\label{unified}
	In the nonparametric implementation I set the mathematical foundations \cite{nonparafuncarxiv} for the completely new robust methodology using the Jordan Decomposition Theorem for a signed measure. Those results showed it to be superior to existing methods with improvements to the parametric version under various settings. However, it still ensured for equivalency to the binomial regression framework that $ \nu^- = 1- F^* = 1 - F $. However, this condition need not hold at all even if $ F^* = F $. To relax this assumption consider a likelihood function as follows,\e \label{uniflik}
	L(X|\b{\beta}) = \b{c}(\b{\nu^+}(\b \lambda))^{(k)}(\b{\nu^-}(\b \lambda))^{(n-k)},\b c \in \mc{R},\ k = \sum_{i=1}^ny_i.
	\ee
	Therefore the link condition holding pointwise takes the following formulation, \begin{eqnarray} 
		(\b{\nu^+})^{\alpha_1^*} = \b \lambda_{|S^+}(X, \b\beta),\\
		(\b{\nu^-})^{\alpha_2^*} = \b \lambda_{|S^-}(X, \b\beta).
	\end{eqnarray}
	
	This general framework then requires a substantially more intricate proof to guarantee almost sure convergence through Latent Adaptive Hierarchical EM Like (LAHEML) algorithm \citep{sdss21} extended to all measure spaces whether finite or $ \sigma- $finite. To see this the theorems below make this formulation clear in a mathematically rigorous way. In doing so, it adds to some well-known results in Real Analysis and Pure Mathematics.
	
	\begin{theorem}\label{theo4}
		Let X be a locally compact and Hausdorff topological space such that $ (X, \Sigma, \nu) $ is a $ \sigma- $finite measure space. Then under Latent Adaptive Hierarchical EM Like Formulation,\e \hat{\beta} \xrightarrow[\text{}]{\text{a.s.}} \beta \ee in $ L^p(X, \nu) $, where $ p \in \{1,...,\infty\}, $ where $ p = \infty $ represents the essentially bounded case. 
	\end{theorem}

	\begin{remark} 
		Therefore, the parameter estimates in any GLM may be estimated almost surely as M-estimators over fintie or $ \sigma- $finite measure spaces. 
	\end{remark}
	
	\begin{remark}
		The formulation above is unqiue and the operator used ensures that convergence can occur uniquely without the need for independence or identical distribution assumption.
	\end{remark}
	
	\begin{remark}
		If the independent and identically distributed results hold then it is but one formulation out of the infinitely many possible formulations, and the results attained hold for this case as well.
	\end{remark}
	
	\begin{remark}
		No assumption is required here other than a Hausdorff topology on compact topological spaces. Any measure that maps a sample space to the reals is shown to have this property locally.
	\end{remark}
	
	As such the results present the only known methodology to the author that allows such flexibility under minimal assumptions of a Hausdorff topology over compact spaces. Thus, it remains to see its performance on simulations and real world datasets. These results are discussed below.
	
	\section{Monte Carlo Simulations}\label{simulations}
	
	In order to validate the robustness of the proposed unified methodology a Bayesian framework is used for extensive simulation studies, penalized and unpenalized, on various DGP's, both symmetric (Logit and Probit) and asymmetric (Complementary Log-Log). For this purpose, datasets were generated from the standard normal distribution as before, for different sample sizes ($ n=\{100,500,1000, 2000\} $) and models,\begin{eqnarray}
		\b{y} = Intercept + X_1 + X_2,\\
		\b{y} = Intercept + X_1 + exp(X_2),\\
		\b{y} = Intercept + exp(X_1) + sin(X_2).
	\end{eqnarray}
	All datasets as before, had 3 parameters to estimate, for the intercept ($ \beta_1 $) and for two explanatory or independent variables drawn from the standard normal ($ \{\beta_2, \beta_3\} $) with the appropriate transformations indicated above. Then for known $ \beta $ values, a Probit, Logit or a Complementary Log-Log DGP was used to generate outcomes (dependent variable $ \b{y} $), that varied in the number of successes present. 
	
	In particular, the known $ \{X,\b\beta\}  $ values along with each functional form above can be used to calculate the probability of each observation for each specific model. Thus, we can consider the calculated $\b{y}$ values along with the generated $X$'s as the data on which we can fit our chosen statistical models for each DGP. Finally, another step was done to create datasets which had different numbers of successes as opposed to failures. Thus, the unbalancedness of the data were varied between $ \{0.1, 0.2, 0.3, 0.4, 0.5\} $. Here, 0.5 indicates equal number of successes and failures (balanced), 0.4 indicates 10\% fewer successes than failures and so forth. Thus, for each model, there are 60 different datasets, each with 3 parameters to estimate, for a total of $ 180\times 3 = 540 $ parameters to estimate, compare and contrast. The penalized results are summarized below.
	
	\begin{table}[!h]
		\caption{Simulation Coverage in Percentage Summary All DGPs (at 1\% Significance Level, Reported in Percentage)}
		\begin{center}
				\begin{tabular}{cccccccc}\hline\hline
					& Prop. & Prop. &Prop. &Prop. & Prop.& Bayesian &Pen. \\
					& N& NU&NP&NPU&Log.& Probit & Logit\\
					\hline
					LGR Covr. (NL)&97.37&100.00&98.68&98.68&97.37&77.63&68.42\\
					PR Covr. (NL)&98.33&98.33&96.67&98.33&95.00&70.00&53.33\\
					Comp. Lg. Covr. (NL)&98.75&100.00&98.75&100.00&100.00&88.75&72.50\\
					LGR Covr. (Mx.)&100.00&100.00&100.00&100.00&100.00&83.33&75.00\\
					PR Covr. (Mx.)&100.00&100.00&100.00&98.68&100.00&89.47&69.74\\
					Comp. Lg. Covr. (Mx.)&98.75&100.00&98.75&98.75&98.75&90.00&69.74\\
					LGR Covr. (L)& 98.68  & 98.68  & 100.00 & 98.68  & 100.00  & 81.58 &73.68\\
					PR Covr. (L)& 97.50 & 97.50 & 98.75 &  96.25 & 95.00  & 88.75 &75.00\\
					Comp. Lg. Covr. (L)&  100.00 & 100.00 &  100.00  & 98.75  & 100.00 & 86.25 & 73.75\\
					\hline
				\end{tabular}
		\end{center}
		\label{simsum1c4}
		\scriptsize{Note: Prop. N., indicates Proposed Nonpenalized, Prop. NU., indicates Proposed Nonpenalized Unified method, Prop. NP., indicates Proposed Penalized, Prop. NPU., indicates Proposed Penalized Unified method, Prop. Log., indicates Parametric Logistic of \cite{chow21}, Bayesian Probit indicates the Bayesian Latent Probit, and Pen. Logit indicates the maximum likelihood Penalized Logistic regression. This is a summary over all three DGPs (Logistic, Probit and Complementary Log-Log), run over sample sizes of n = \{100, 500, 1000, 2000\} and unbalancedness of \{0.1, 0.2, 0.3, 0.4, 0.5\} for all linear, non-linear and mixed models fitted (here 0.5 indicates equal number of 1's and 0's (balanced), 0.4 indicates 10\% fewer 1's than 0's and so forth). For each DGP there are 20 different datasets to consider for each of the linear, mixed and non-linear models considered for a total of 60 different datasets per DGP. For each dataset there are three parameters of interest or $ \beta $'s. In total there are 180 parameters per DGP for a total of 540 parameters to be estimated over the entire simulation study. The results are summarized by average over all simulated datasets.} 
	\end{table}
	\begin{table}[!ht]
		\caption{Simulation Confidence Interval Range for All DGPs (at 1\% Significance Level)}
		\begin{center}
				\begin{tabular}{cccccccc}\hline\hline
					& Prop. & Prop. &Prop. &Prop. & Prop.& Bayesian &Pen. \\
					& N& NU&NP&NPU&Log.& Probit & Logit\\
					\hline
					LGR Covr. (NL)&6.01&5.76&5.85&5.43&5.58&1.79&5.40\\
					PR Covr. (NL)&5.92&6.10&5.46&5.60&5.15&2.07&5.22\\
					Comp. Lg. Covr. (NL)&5.64&5.69&5.53&5.29&5.58&1.90&5.58\\
					LGR Covr. (Mx.)&6.47&5.67&5.37&6.19&5.67&2.20&6.52\\
					PR Covr. (Mx.)&5.95&5.75&5.29&5.46&5.40&2.13&6.01\\
					Comp. Lg. Covr. (Mx.)&6.1&5.46&5.74&5.74&5.34&2.03&5.24\\
					LGR Covr. (L)& 5.71  & 5.65 & 5.67 & 5.73  &  5.61 & 1.86 &6.13\\
					PR Covr. (L)&  5.74 & 5.43 & 5.49 &  5.60 &  5.42 & 2.01 &6.45\\
					Comp. Lg. Covr. (L)&  5.92 &  5.58 & 5.60  & 5.49  &  5.36 & 1.87 &6.25\\
					\hline
				\end{tabular}
		\end{center}
		\label{simsum2c4}
		\scriptsize{Note: Prop. N., indicates Proposed Nonpenalized, Prop. NU., indicates Proposed Nonpenalized Unified method, Prop. NP., indicates Proposed Penalized, Prop. NPU., indicates Proposed Penalized Unified method, Prop. Log., indicates Parametric Logistic of \cite{chow21}, Bayesian Probit indicates the Bayesian Latent Probit, and Pen. Logit indicates the maximum likelihood Penalized Logistic regression. This is a summary over all three DGPs (Logistic, Probit and Complementary Log-Log), run over sample sizes of n = \{100, 500, 1000, 2000\} and unbalancedness of \{0.1, 0.2, 0.3, 0.4, 0.5\} for all linear, non-linear and mixed models fitted (here 0.5 indicates equal number of 1's and 0's (balanced), 0.4 indicates 10\% fewer 1's than 0's and so forth). For each DGP there are 20 different datasets to consider for each of the linear, mixed and non-linear models considered for a total of 60 different datasets per DGP. For each dataset there are three parameters of interest or $ \beta $'s. In total there are 180 parameters per DGP for a total of 540 parameters to be estimated over the entire simulation study. The results are summarized by average over all simulated datasets.} 
	\end{table}
	\begin{table}[h]
		\caption{Simulation Summary of ARS for All DGPs}
		\begin{center}
				\begin{tabular}{cccccccc}\hline\hline
					& Prop. & Prop. &Prop. &Prop. & Prop.& Bayesian &Neural \\
					& N& NU&NP&NPU&Log.& Probit & Net\\
					\hline
					Non-Linear&0.07  &0.10  &0.07  &0.05   &0.23   &0.25  &0.16\\
					Mixed&0.17   &0.05  &0.07  &0.07   &0.23   &0.27    &0.21\\
					Linear& 0.08  &0.07  &0.05  & 0.05  & 0.19  & 0.23   &0.21\\
					
					\hline
				\end{tabular}
		\end{center}
		\label{simsum3c4}
		\scriptsize{Note: Prop. N., indicates Proposed Nonpenalized, Prop. NU., indicates Proposed Nonpenalized Unified method, Prop. NP., indicates Proposed Penalized, Prop. NPU., indicates Proposed Penalized Unified method, Prop. Log., indicates Proposed Logistic, Bayesian Probit indicates the Bayesian Latent Probit, and Pen. Logit indicates the maximimum likelihood Penalized Logistic regression. This is a summary over all three DGPs (Logistic, Probit and Complementary Log-Log), run over sample sizes of n = \{100, 500, 1000, 2000\} and unbalancedness of \{0.1, 0.2, 0.3, 0.4, 0.5\} for all linear, non-linear and mixed models fitted (here 0.5 indicates equal number of 1's and 0's (balanced), 0.4 indicates 10\% fewer 1's than 0's and so forth). For each DGP there are 20 different datasets to consider for each of the linear, mixed and non-linear models considered for a total of 60 different datasets per DGP. For each dataset there are three parameters of interest or $ \beta $'s. In total there are 180 parameters per DGP for a total of 540 parameters to be estimated over the entire simulation study. The results are summarized by average over all simulated datasets.} 
	\end{table}
	%
	\section{Empirical Application}\label{application}
	
	To compare and contrast the methodology with those presented in the previous chapters, I apply it to various datasets below. The datasets include the Intoxication dataset as well as the Higgs dataset used in \cite{sdss21}, in addition to the Challenger space shuttle disaster dataset of 1986. Below more detailed explanations are given for these datasets avoiding duplication where possible.
	
	\subsection{Detecting Heavy Drinking Events Using Smartphone Data}
	
	To detect heavy drinking events using smartphone accelerometer data in \cite{killian2019learning}, I run the unified methodology in a penalized and unpenalized setting. For completeness note that the authors identified heavy drinking events within a four second window of their measured variable of Transdermal Alcohol Content (TAC) on smartphone accelerometer data. Their best classifier was a Random Forest with about 77.50\% accuracy. A similar analysis was done on a far simpler model of TAC readings against the accelerometer readings as predictors, for all subject's phone placement in 3D space, for the x, y and z axes, \e TAC = Intercept + x-axis\ reading + y-axis\ reading + z-axis\ reading.\ee TAC here was set to 1 if the measurement was over 0.08 and 0 otherwise. The same four second time window of accelerometer readings was used in the analysis with the assumption that the TAC readings were unlikely to change in such a small time interval. An application of the current methodology showed it to be extremely effective with only $ 1,000 $ iterations.
	
\begin{figure}[!htb]
		\begin{minipage}{0.5\textwidth}
			\centering
			\includegraphics[width=0.92\linewidth]{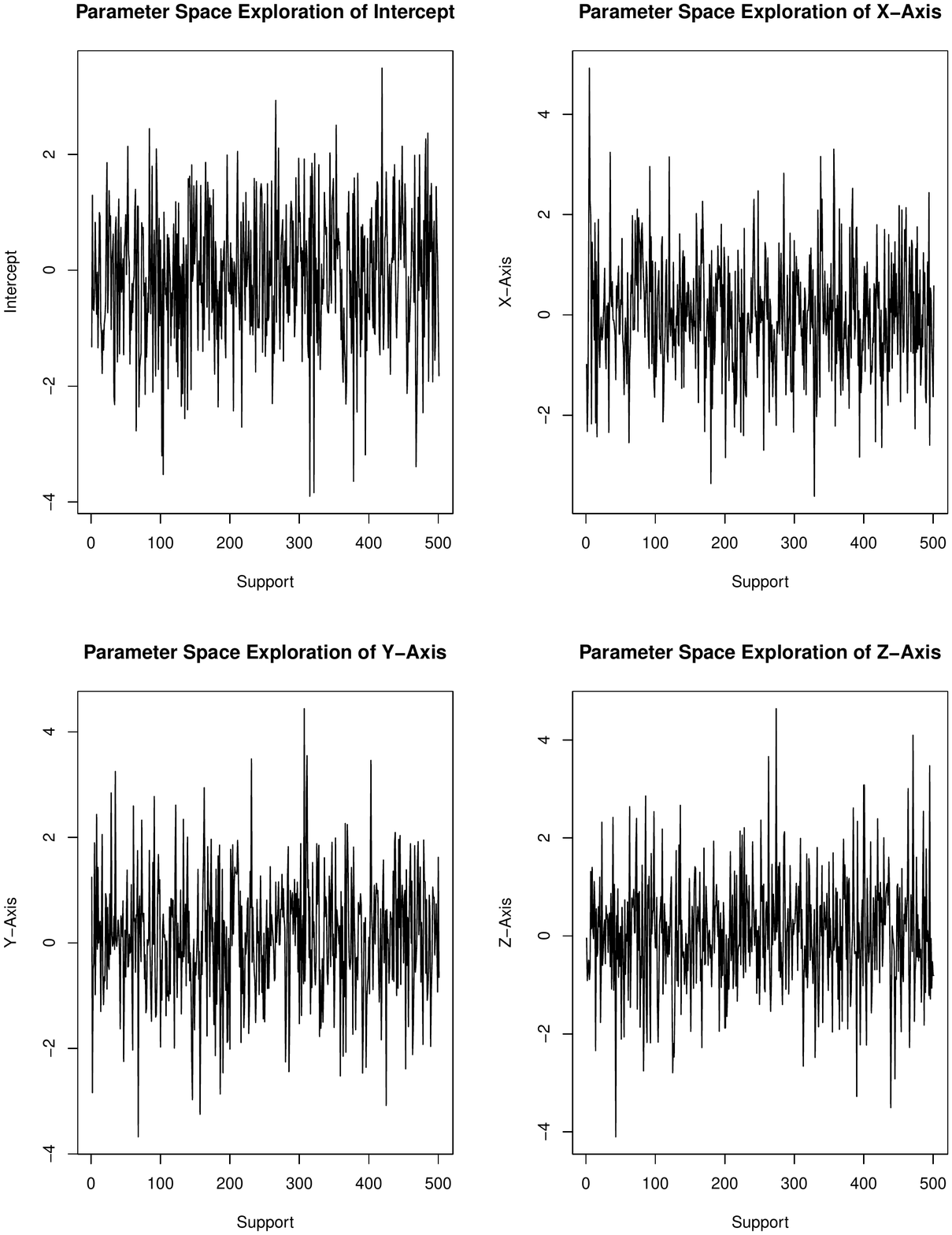}
			\caption{Heavy Drinking Event Data Sample Space Exploration Plot for Nonparametric Unified Methodology.}
			\label{fig:intxunifconv}
		\end{minipage}%
		\begin{minipage}{0.5\textwidth}
			\centering
			\includegraphics[width=0.92\linewidth]{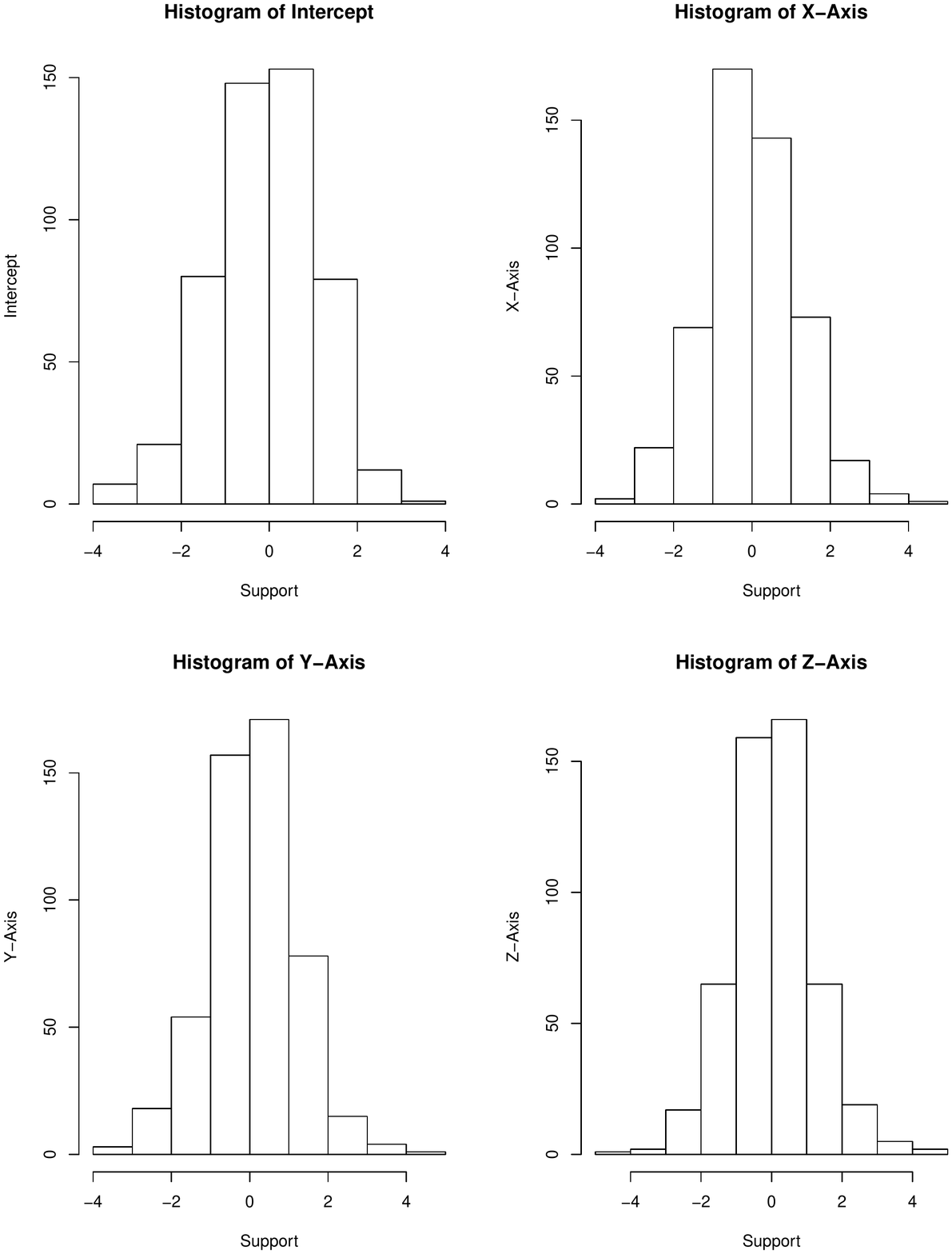}
			\caption{Heavy Drinking Event Data  Histogram of Parameters for Nonparametric Unified Methodology.}
			\label{fig:intxunifhist}
		\end{minipage}
	\end{figure}

\begin{figure}[!htb]
		\begin{minipage}{0.5\textwidth}
			\centering
			\includegraphics[width=0.92\linewidth]{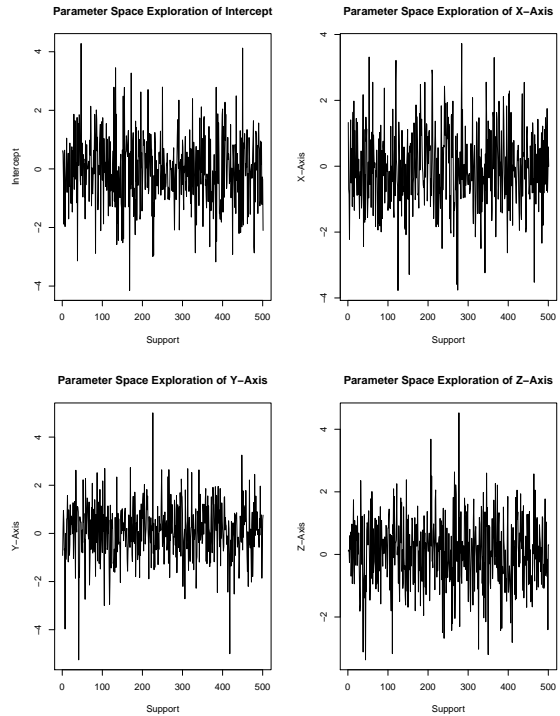}
			\caption{Heavy Drinking Event Data  Sample Space Exploration Plot for Nonparametric Penalized Unified Methodology.}
			\label{fig:intxpenunifconv}
		\end{minipage}%
		\begin{minipage}{0.5\textwidth}
			\centering
			\includegraphics[width=0.92\linewidth]{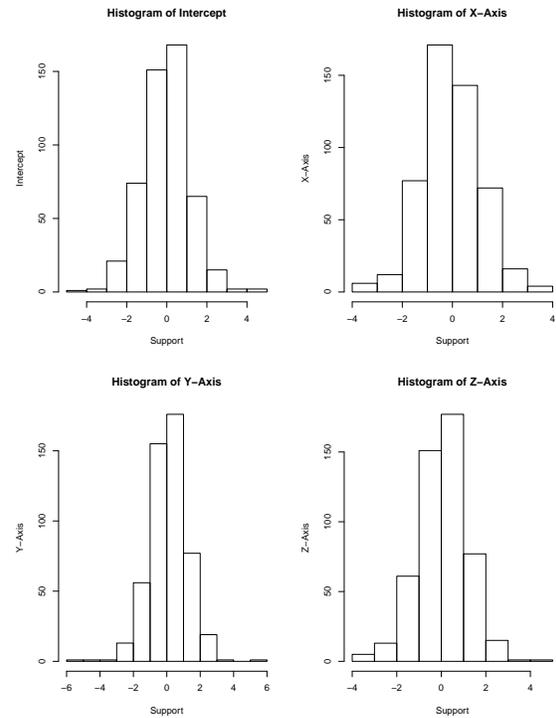}
			\caption{Heavy Drinking Event Data  Histogram of Parameters for Nonparametric Penalized Unified Methodology.}
			\label{fig:intxpenunifhist}
		\end{minipage}
	\end{figure}

\begin{figure}[!htb]
		\begin{minipage}{0.5\textwidth}
			\centering
			\includegraphics[width=0.8\linewidth]{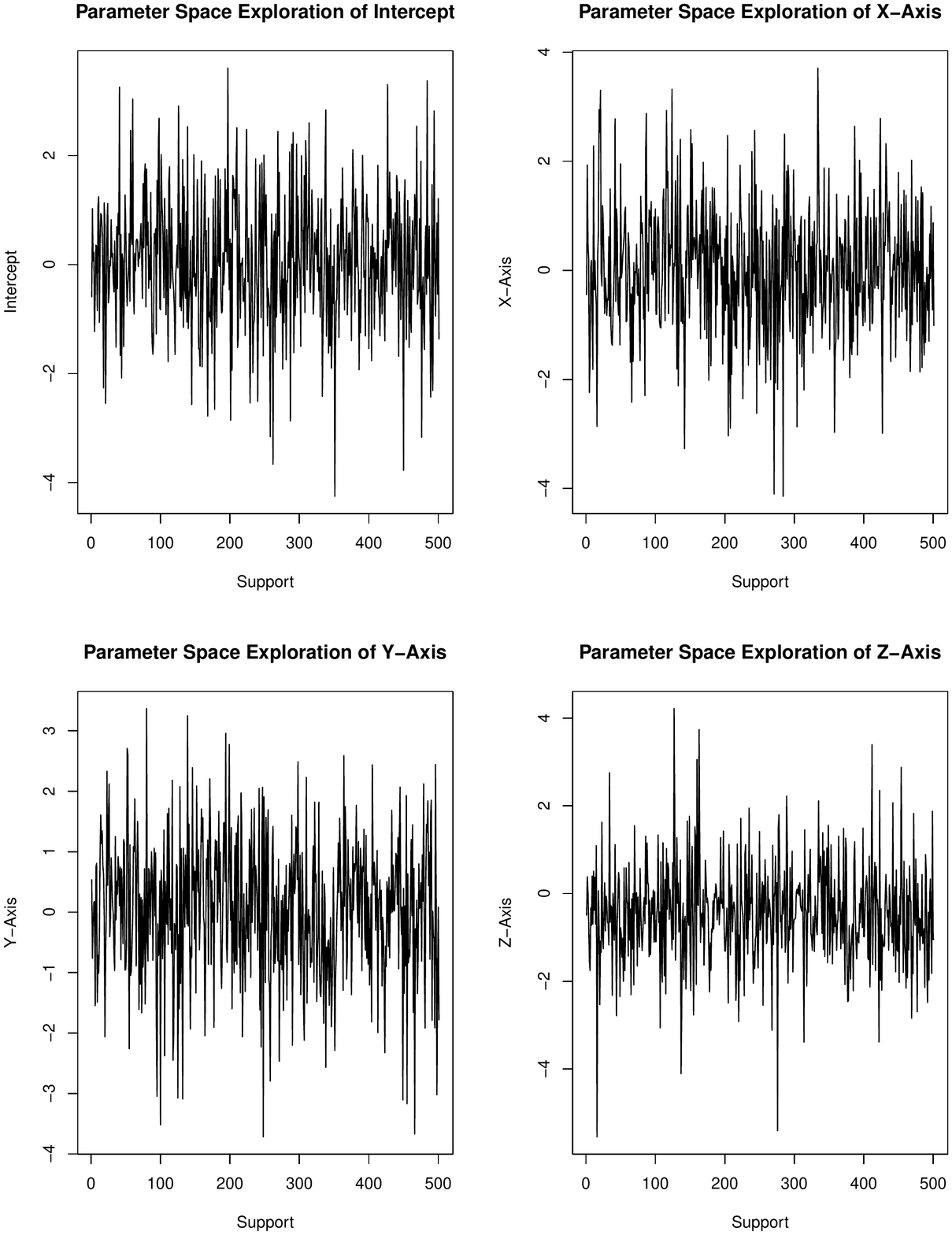}
			\caption{Heavy Drinking Event Data  Sample Space Exploration Plot for Nonparametric Methodology.}
			\label{fig:intxconv2}
		\end{minipage}%
		\begin{minipage}{0.5\textwidth}
			\centering
			\includegraphics[width=0.8\linewidth]{IntxUnifHist}
			\caption{Heavy Drinking Event Data  Histogram of Parameters for Nonparametric Methodology.}
			\label{fig:intxhist2}
		\end{minipage}
	\end{figure}

\begin{figure}[!htb]
		\begin{minipage}{0.5\textwidth}
			\centering
			\includegraphics[width=0.92\linewidth]{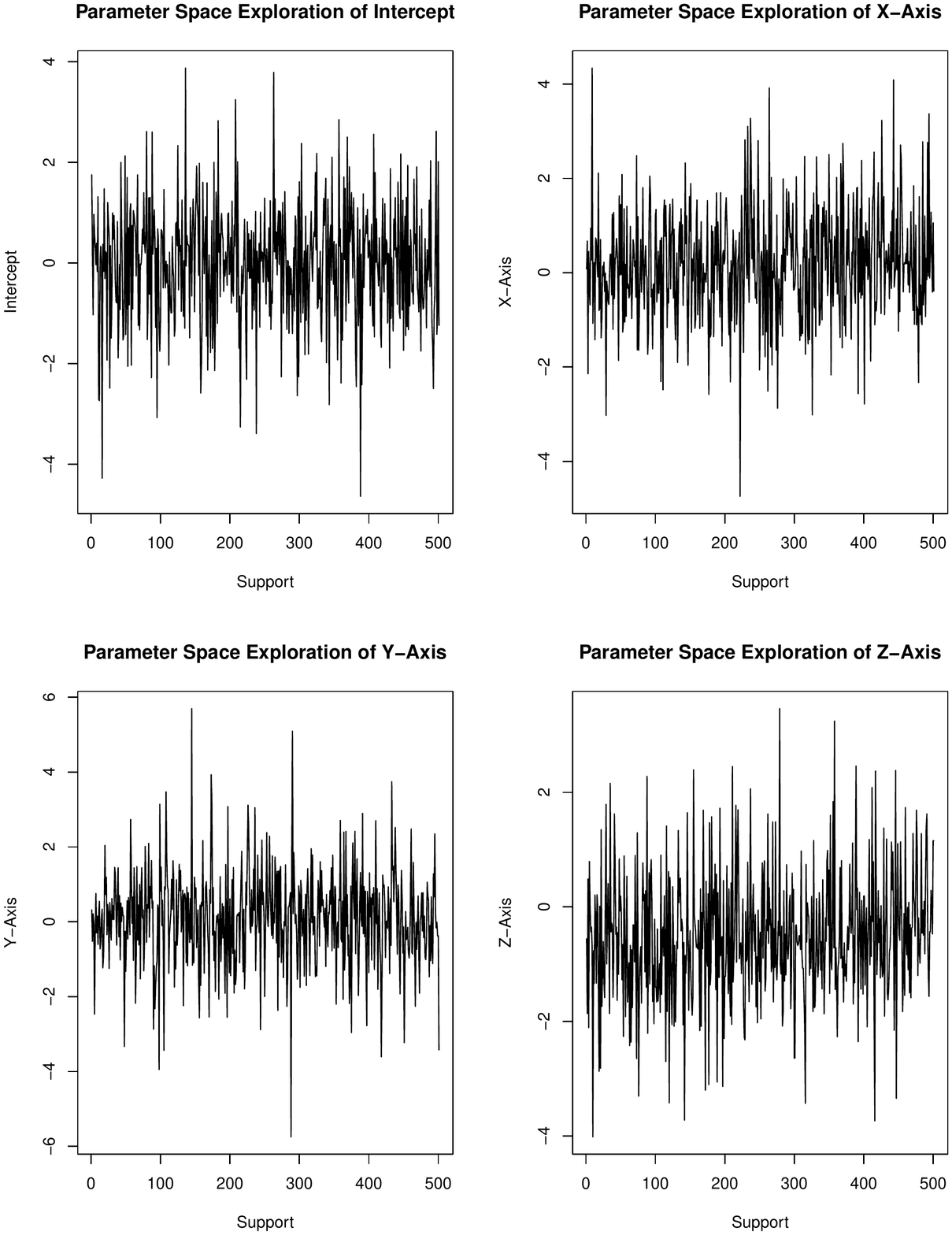}
			\caption{Heavy Drinking Event Data  Sample Space Exploration Plot for Nonparametric Penalized Methodology.}
			\label{fig:intxpenconv2}
		\end{minipage}%
		\begin{minipage}{0.5\textwidth}
			\centering
			\includegraphics[width=0.92\linewidth]{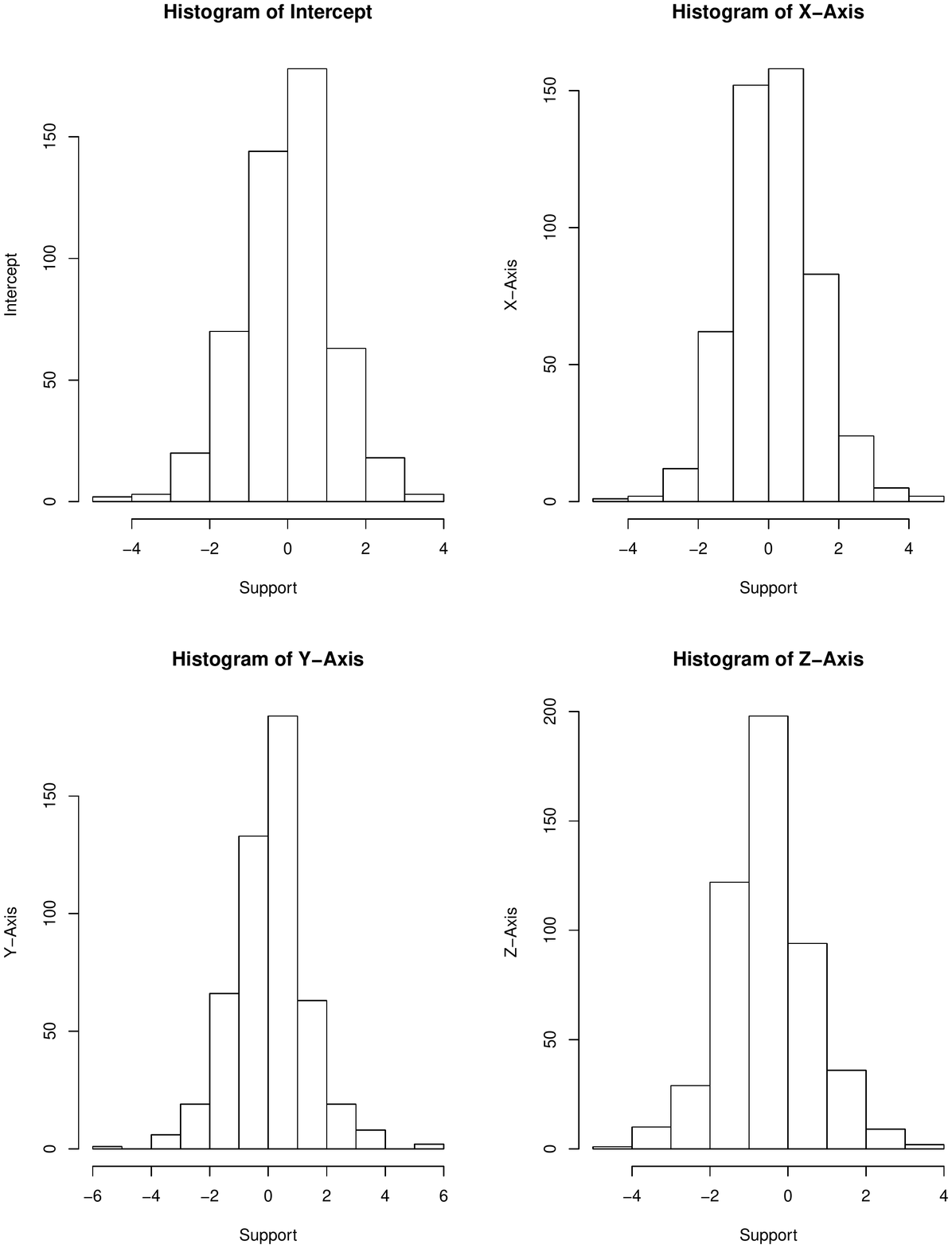}
			\caption{Heavy Drinking Event Data  Histogram of Parameters for Nonparametric Penalized Methodology.}
			\label{fig:intxpenhist2}
		\end{minipage}
	\end{figure}

	In particular, in the current application all methodologies were run for $ 1,000 $ iterations and below I present the convergence and histogram plots in Figure \ref{fig:intxunifconv}, Figure \ref{fig:intxunifhist}, Figure \ref{fig:intxpenunifconv}, Figure \ref{fig:intxpenunifhist}, Figure \ref{fig:intxconv2}, Figure \ref{fig:intxhist2}, Figure \ref{fig:intxpenconv2}, and Figure \ref{fig:intxpenhist2} for both the unified and nonunified methodologies.
	The results tell a fascinating story regarding MIPs. The Test Data (TeD) classification results for the nonunified methodology were excellent, with perfect identification. For the Training (TrD), however, the results were less effective with around $ 70.00\% $ accuracy. This trend though was largely consistent with the results from the unified methodology also giving perfect classification in TeD with the TrD classification results slightly worse, though not significantly so.
	\begin{table}[h!]
		\begin{minipage}{0.5\textwidth}
			\caption{Intoxication Dataset Summary of ARS for All Relevant Methodologies}
			\begin{center}
					\begin{tabular}{ccc}\hline\hline
						Methodology&TrD&TeD\\
						\hline
						Unified Penalized  & 0.21 & 0.06 \\
						Penalized Nonparametric & 0.34 & 0.00 \\
						Nonparametric & 0.30 & 0.00 \\
						Unified Nonparametric & 0.24 & 0.00 \\
						Parametric & 0.68 & 0.76 \\
						Existing Bayes & 0.77 & 0.66 \\
						MLE Logistic & 0.91 & 0.90 \\
						Penalized Logistic & 0.91 & 0.90 \\
						\hline
					\end{tabular}
			\end{center}
			\label{IntxARS}
			\scriptsize{} 
		\end{minipage}
		\begin{minipage}{0.5\textwidth}
			\caption{Intoxication Dataset Summary of AIC for All Relevant Methodologies}
			\begin{center}
					\begin{tabular}{ccc}\hline\hline
						Methodology&TrD&TeD\\
						\hline
						Unified Penalized & 1.32 & 0.22 \\
						Penalized Nonparametric& 3.16 & 0.95 \\
						Nonparametric & 3.13 & 0.94 \\
						Unified Nonparametric& 1.77 & 1.07 \\
						Parametric & 1.31 & 0.97 \\
						Existing Bayes & 1.83 & 1.18 \\
						MLE Logistic & 1.21 & 1.00 \\
						Penalized Logistic & 1.21 & 1.00 \\
						\hline
					\end{tabular}
			\end{center}
			\label{IntxAIC}
			\scriptsize{} 
		\end{minipage}
	\end{table}
	\begin{table}[!h]
		\begin{minipage}{1\textwidth}
			\caption{Intoxication Dataset Parameter Summary for All Relevant Methodologies}
			\begin{center}
					\begin{tabular}{ccccc}\hline\hline
						Predictor&Estimates&CI-Low&CI-High&Methodology\\
						\hline
						Intercept & 0.24** & 0.01 & 0.47 & (1) \\
						X-axis & 0.02 & -0.22 & 0.25 & (1) \\
						Y-axis & 0.03 & -0.19 & 0.25  & (1) \\
						Z-axis & -0.54** & -0.83 & -0.26 & (1) \\
						Intercept & -0.06** & -0.11 & -0.01  & (2) \\
						X-axis & -0.01 & -0.05 & 0.04  & (2) \\
						Y-axis & 0.17** & 0.13 & 0.21  & (2) \\
						Z-axis & -0.08** & -0.13 & -0.02  & (2) \\
						Intercept & 0.22 & -0.03 & 0.48  & (3) \\
						X-axis & -0.07 & -0.31 & 0.17  & (3) \\
						Y-axis & 0.21 & -0.04 & 0.46  & (3) \\
						Z-axis & -0.82** & -1.05 & -0.59  & (3) \\
						Intercept & 0.17 & -0.07 & 0.42 & (4) \\
						X-axis & -0.02 & -0.27 & 0.23  & (4) \\
						Y-axis & 0.16 & -0.06 & 0.38  & (4) \\
						Z-axis & 0.02 & -0.2 & 0.24  & (4) \\
						Intercept & -0.13 & -0.3 & 0.05  & (5) \\
						X-axis & 0.01 & -0.19 & 0.2  & (5) \\
						Y-axis & 0.07 & -0.13 & 0.27  & (5) \\
						Z-axis & -0.21** & -0.37 & -0.05  & (5) \\
						Intercept & -0.01 & -0.14 & 0.11 & (6) \\
						X-axis & -0.12 & -0.32 & 0.08  & (6) \\
						Y-axis & 0.24** & 0.06 & 0.43  & (6) \\
						Z-axis & -0.02 & -0.15 & 0.1  & (6) \\
						Intercept & -0.87*** & -0.9 & -0.85  & (7) \\
						X-axis & -0.04* & -0.09 & 0  & (7) \\
						Y-axis & 0.17*** & 0.11 & 0.23 & (7) \\
						Z-axis & 0.00*** & 0.00 & 0.00  & (7) \\
						Intercept & -0.87*** & -0.9 & -0.84 & (8) \\
						X-axis & -0.04* & -0.11 & 0.02 & (8) \\
						Y-axis & 0.17*** & 0.09 & 0.25  & (8) \\
						Z-axis & 0.00*** & 0.00 & 0.00  & (8) \\					
						\hline
					\end{tabular}
			\end{center}
			\label{IntxParam}
			\scriptsize{Note: (1) Nonprametric, (2) Unified Nonparametric, (3) Penalized Nonparametric, (4) Unified Penalized Nonparametric, (5) Parametric, (6) Existing Bayesian, (7) MLE Logistic, (8) Penalized Logistic.} 
		\end{minipage}
	\end{table}
	
	In regards to inference, the results were largely identical, thus providing further verification for the applicability of the Likelihood Principle, for both the unified and nonunified nonparametric cases. This is a direct result of the linear operators in both methodologies being continuous, as defined from the sample space to the link subspace. However, the proposed methodology has significant advantages over the nonunified case in terms of model fit. That is considering the median $ \b\beta s, $ with the Binomial Likelihood (LAHEML integrates out the probability of success so the original data likelihood is used here), the unified methodology was nearly 2.4 times better in the TrD and almost 4.3 times as good in the TrD for the Proposed Unified Nonparametric methodology over the Proposed Nonparametric methodology. The results are also consistent for the penalized applications. These results are entirely consistent with the underlying mathematical foundations, and I discuss them more in the forthcoming Section \ref{unifdisc}. Please further note that the goal here is to compare the unified and nonunifed applications and not the other models per se\footnote{The Bayesian Latent Probit was run for $ 5,000 $ iterations whereas the Proposed Parametric Logistic was run for only $ 1,000 $ iterations here.}.
	
	\subsection{Exotic Particle Detection Using Particle Accelerator Data}
	
\begin{figure}[!htb]
		\begin{minipage}{1\textwidth}
			\centering
			\includegraphics[width=0.92\linewidth]{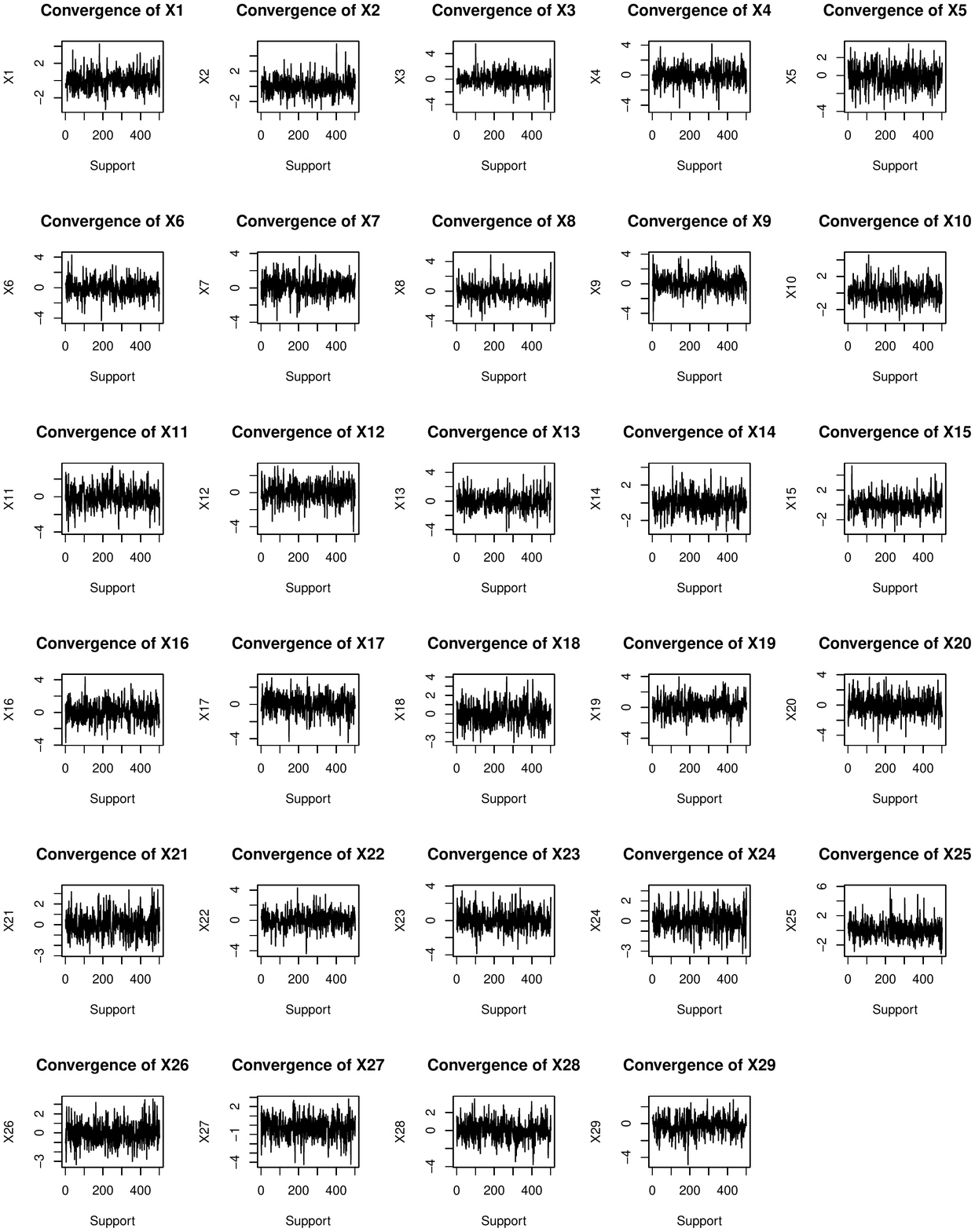}
			\caption{Exotic Particle Detection Data Sample Space Exploration Plot for Nonparametric Unified Methodology.}
			\label{fig:higgsunifconvc4}
		\end{minipage}
	\end{figure}

\begin{figure}
		\begin{minipage}{1\textwidth}
			\centering
			\includegraphics[width=0.92\linewidth]{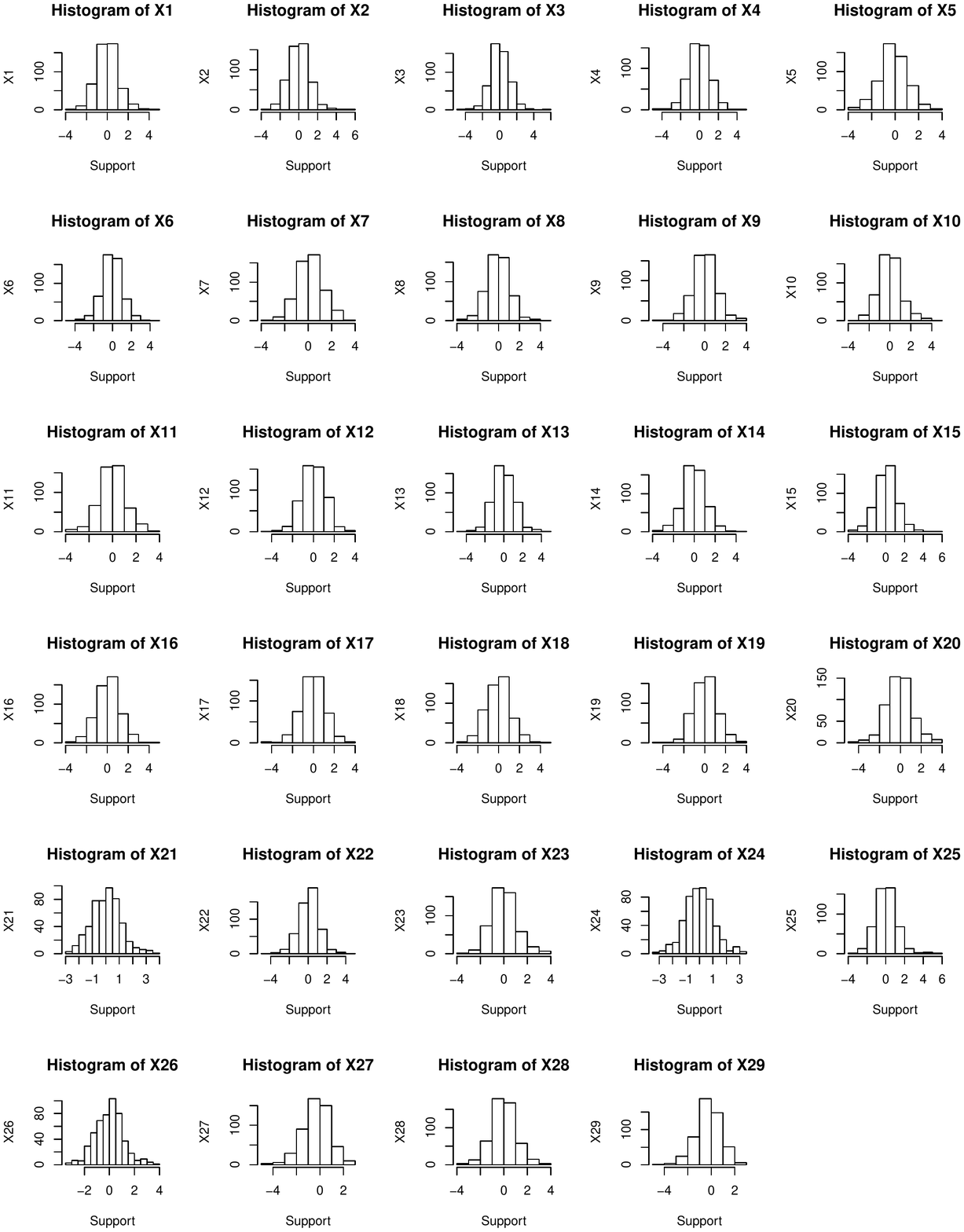}
			\caption{Exotic Particle Detection Data Histogram of Parameters for Nonparametric Unified Methodology.}
			\label{fig:higgsunifhistc4}
		\end{minipage}
	\end{figure}

\begin{figure}[!htb]
		\begin{minipage}{1\textwidth}
			\centering
			\includegraphics[width=0.92\linewidth]{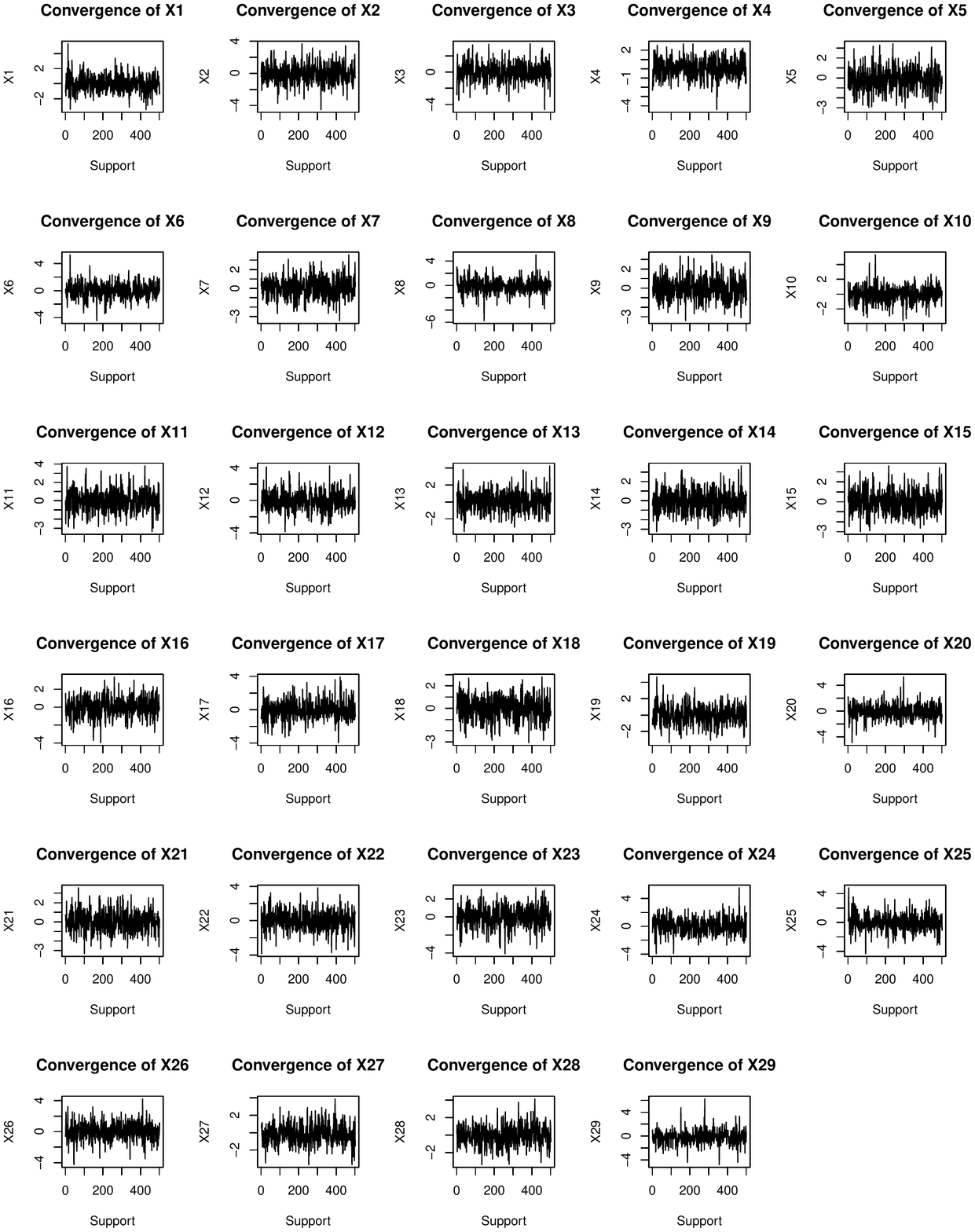}
			\caption{Exotic Particle Detection Data Sample Space Exploration Plot for Nonparametric Penalized Unified Methodology.}
			\label{fig:higgspenunifconvc4}
		\end{minipage}%
	\end{figure}

\begin{figure}
		\begin{minipage}{1\textwidth}
			\centering
			\includegraphics[width=0.92\linewidth]{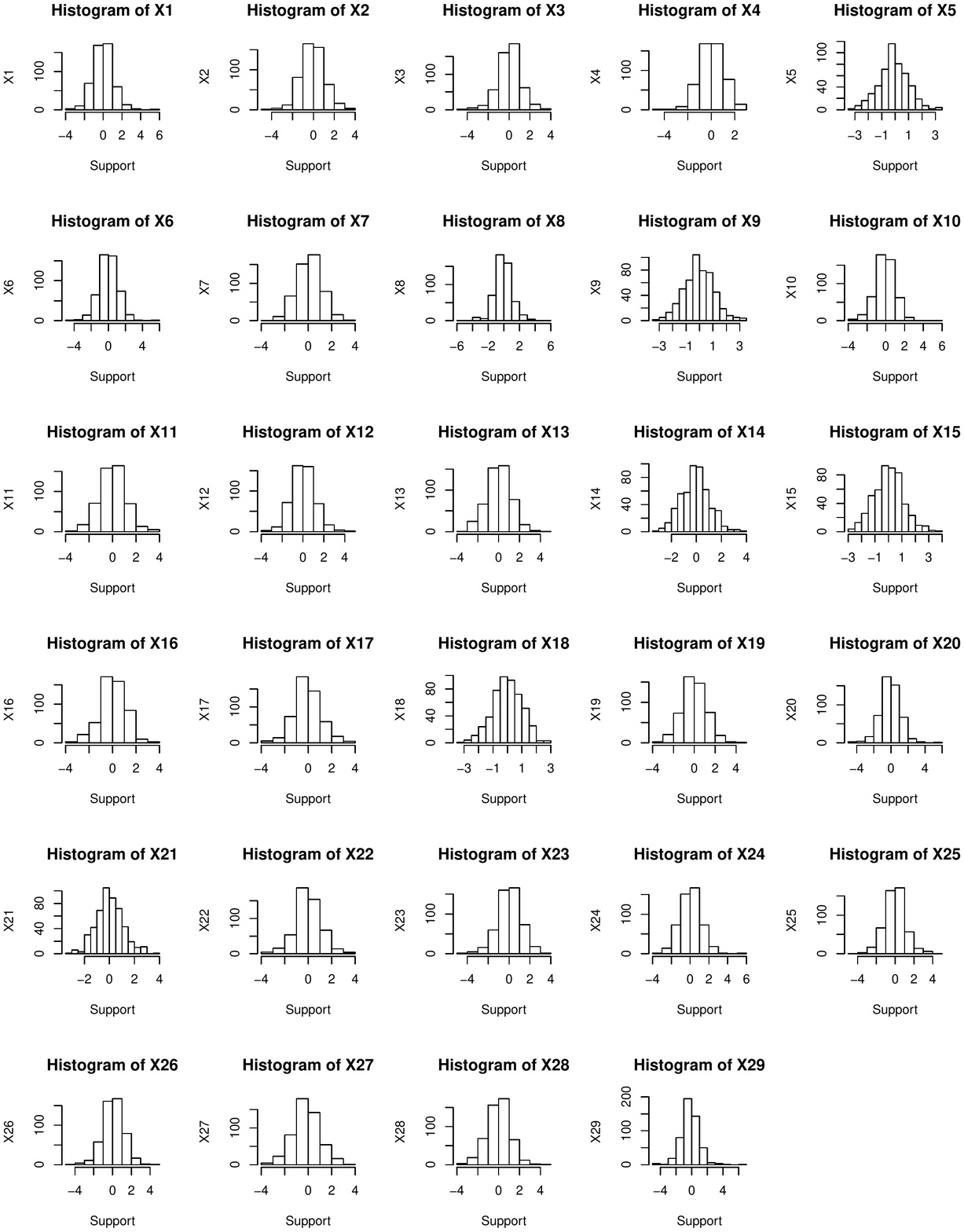}
			\caption{Exotic Particle Detection Data Histogram of Parameters for Nonparametric Penalized Unified Methodology.}
			\label{fig:higgspenunifhistc4}
		\end{minipage}
	\end{figure}

\begin{figure}[!htb]
		\begin{minipage}{1\textwidth}
			\centering
			\includegraphics[width=0.92\linewidth]{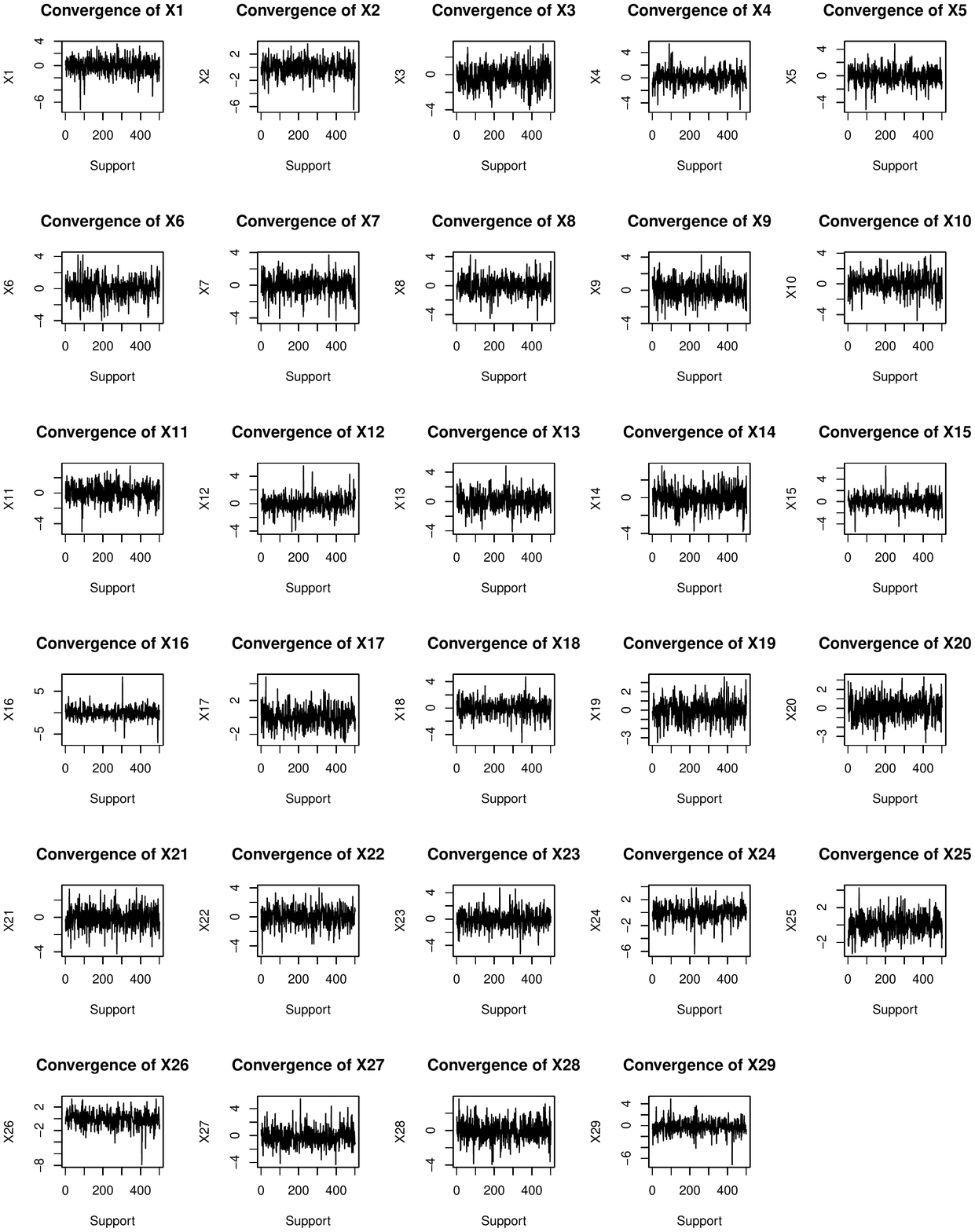}
			\caption{Exotic Particle Detection Data Sample Space Exploration Plot for Nonparametric Methodology.}
			\label{fig:higgsconvc4}
		\end{minipage}%
	\end{figure}

\begin{figure}
		\begin{minipage}{1\textwidth}
			\centering
			\includegraphics[width=0.92\linewidth]{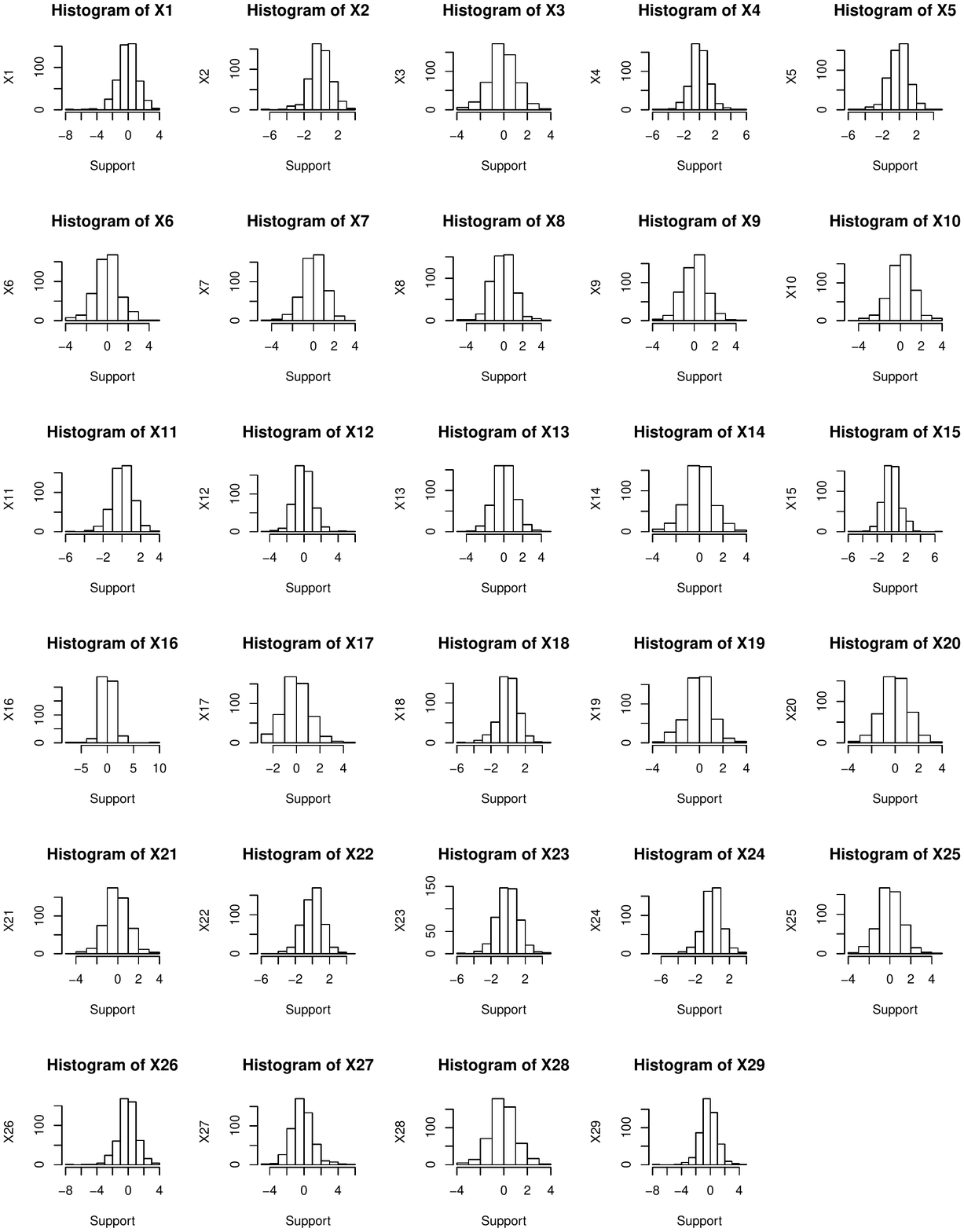}
			\caption{Exotic Particle Detection Data Histogram of Parameters for Nonparametric Methodology.}
			\label{fig:higgshistc4}
		\end{minipage}
	\end{figure}

\begin{figure}[!htb]
		\begin{minipage}{1\textwidth}
			\centering
			\includegraphics[width=0.92\linewidth]{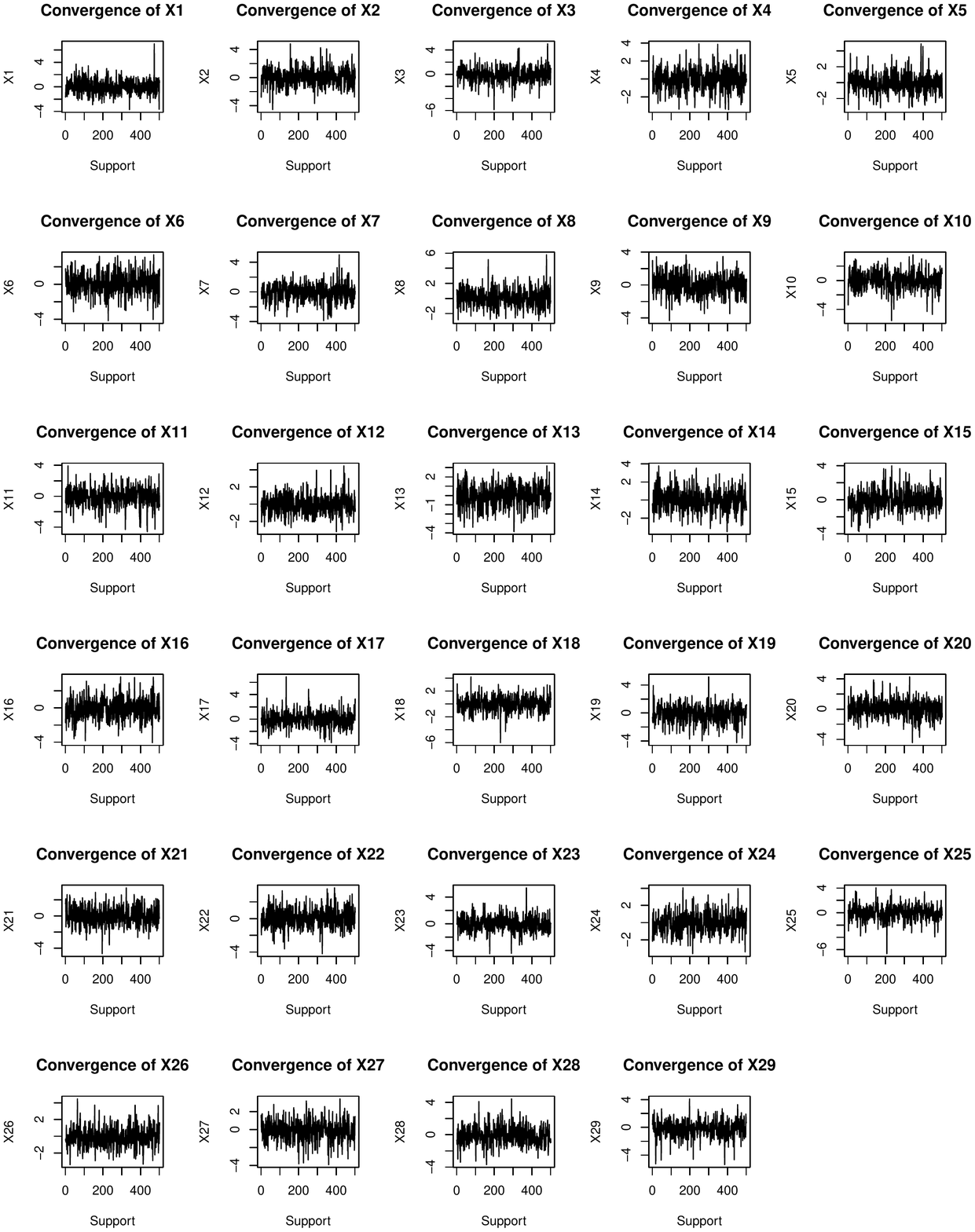}
			\caption{Exotic Particle Detection Data Sample Space Exploration Plot for Nonparametric Penalized Methodology.}
			\label{fig:higgspenconvc4}
		\end{minipage}%
	\end{figure}

\begin{figure}
		\begin{minipage}{1\textwidth}
			\centering
			\includegraphics[width=0.92\linewidth]{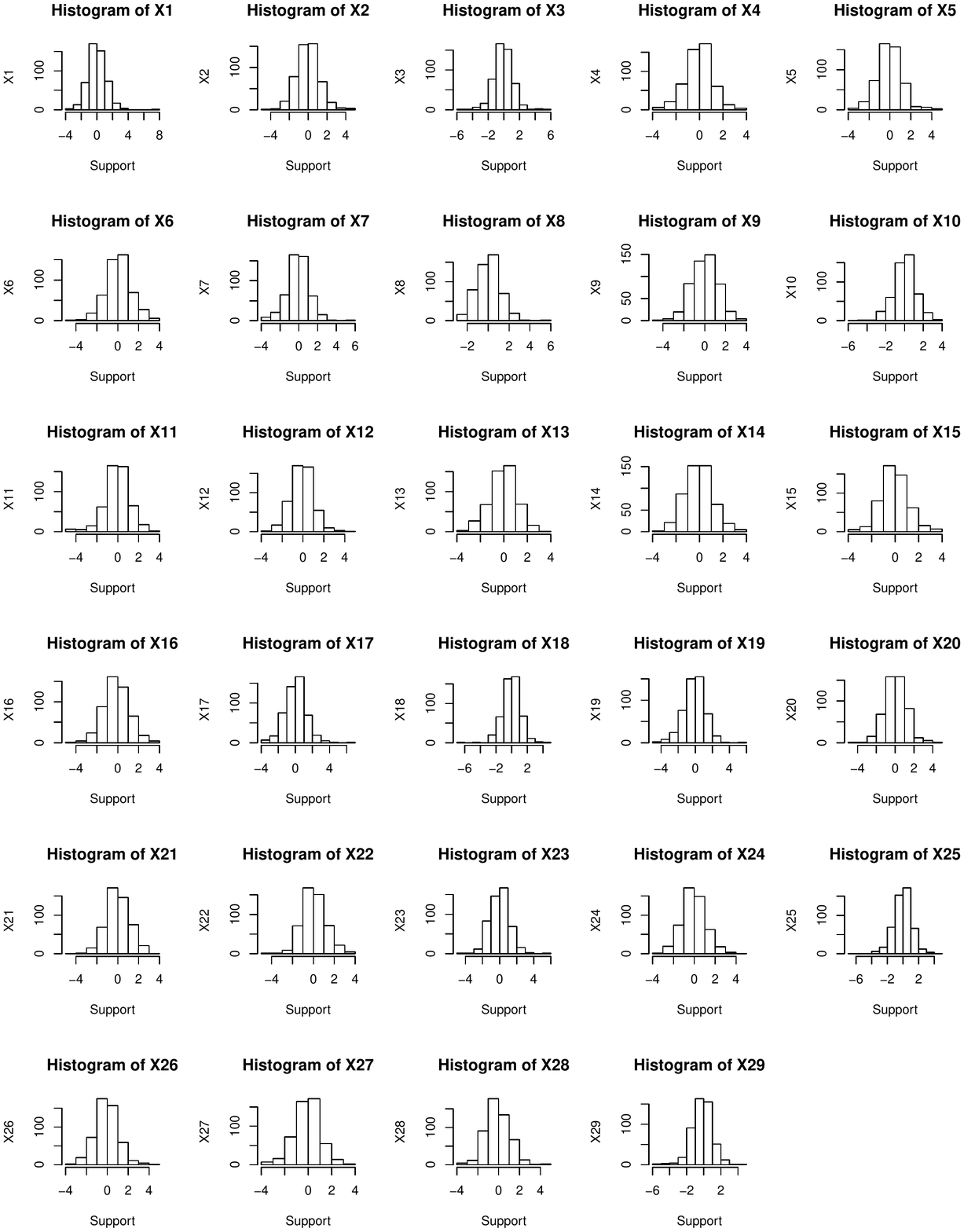}
			\caption{Exotic Particle Detection Data Histogram of Parameters for Nonparametric Penalized Methodology.}
			\label{fig:higgspenhistc4}
		\end{minipage}
	\end{figure}

	The second application of the methodology was for identifying high-energy particles in Physics (\cite{baldi2014searching}). Note that there are 28 feature sets in the paper, with the first 21 features the kinematic properties measured by detectors in particle accelerators. The last 7 high-level features were derived from the first 21 to discriminate between the two classes. Therefore, inference is not the primary purpose for this application. The classes to be identified is either 0 and 1, and refer to noise and signal respectively. For completeness, the model specification is restated below. \e
	Signal/Noise = Intercept + \sum_{i=1}^{28} Feature_i.
	\ee
	For more information on the actual feature sets I refer the reader to the original paper. Given the large datasize, for these applications, LAHEML was run for only $ 1,000 $ iterations with $ 500 $ burn-in period. The convergence plots, along with the histograms of each parameter may be found in Figure \ref{fig:higgsunifconvc4}, Figure \ref{fig:higgsunifhistc4}, Figure \ref{fig:higgspenunifconvc4}, Figure \ref{fig:higgspenunifhistc4},  Figure \ref{fig:higgsconvc4}, Figure \ref{fig:higgshistc4}, Figure \ref{fig:higgspenconvc4} and Figure \ref{fig:higgspenhistc4}. The penalized and unpenalized estimation formulations were identical to that for the Intoxication application for Biostatistics. The classification outcomes were extremely encouraging, and can be found in Table \ref{HiggsARS4}. The AICs can be found in Table \ref{HiggsAIC4}.
	\begin{table}[!h]
		\begin{minipage}{0.5\textwidth}
			\caption{Higgs Dataset Parameter Summary for All Relevant Methodologies}
			\begin{center}
					\begin{tabular}{ccc}\hline\hline
						Methodology&TrD&TeD\\
						\hline
						Unified Penalized  & 0.41 & 0.18 \\
						Penalized Nonparametric & 0.48 & 0.19 \\
						Nonparametric & 0.51 & 0.19\\
						Unified Nonparametric & 0.53 & 0.16 \\
						Parametric & 0.77 & 0.77\\
						Existing Bayes & 0.67 & 0.67 \\
						MLE Logistic & 0.58 & 0.58 \\
						\hline
					\end{tabular}
			\end{center}
			\label{HiggsARS4}
			\scriptsize{} 
		\end{minipage}
		\begin{minipage}{0.5\textwidth}
			\caption{Higgs Dataset Summary of AIC for All Relevant Methodologies}
			\begin{center}
					\begin{tabular}{ccc}\hline\hline
						Methodology&TrD&TeD\\
						\hline
						Unified Penalized & 3.68 & 1.48 \\
						Penalized Nonparametric& 4.11 & 1.65 \\
						Nonparametric & 3.83 & 1.57 \\
						Unified Nonparametric& 3.51 & 1.36 \\
						Parametric & 1.36 & 1.36 \\
						Existing Bayes & 0.83 & 0.75 \\
						MLE Logistic & 1.28 & 1.28 \\
						\hline
					\end{tabular}
			\end{center}
			\label{HiggsAIC4}
			\scriptsize{} 
		\end{minipage}
	\end{table}
	
	
	
	
	
	Given the small number of iterations one would expect that these MIP results can be further improved for the unified applications for penalized or unpenalized cases. However, even with this small number of iterations, the Unified Penalized and Unified Nonparametric methodologies outperformed their nounified counterparts in TeDs for both ARS and AIC. Thus, it seems reasonable to surmize that with the same number of iterations the unified methodologies would outperform the results of the nonunified applications\footnote{The Bayesian Latent Probit was run for $ 5,000 $ iterations, while the Proposed Parametric application was run for only $ 1,000 $ iterations.}. The goal here is not to compare the performances of the published parametric versions, but rather the Proposed Unified Nonparametric and Proposed Nonparametric versions in both the penalized and unpenalized applications. As such, I discuss these findings along with their mathematical implications more broadly in Section \ref{unifdisc}.
	
	\subsection{Challenger Disaster}
	The final application of the methodology is for the Challenger disaster. This dataset holds a special place in my memory, as it was the first dataset on which I saw categorical models, applied. In fact, that model was the Logistic, which the published and cited works befor,e and this one as well, improves in regads to MIPs. Thus, it seems fitting to have an application on this dataset that started my initial curiosity into such models. To give some background information, note that the Challenger Space Shuttle explosion occured in 1986 due to the failure of an O-ring, a component on the rocket. It is now widely recognized that the material this component was made of, is susceptible to stress especially when the outside temperature is low. There are certain engineering reasons for this, which are not important for the current discussion presently, and I refer the reader to \cite{draper1995assessment} for further discussions on this. In addition, unlike in other papers that use the same dataset (see for example \cite{draper1995assessment}) I change the task of interest slightly here, from understanding the number of O-rings under thermal stress to understanding the probability of an O-ring experiencing thermal stress as a function of the outside temperature and a variable called leak-check pressure. Thus, the model may be given as, \e O-ring\ Under\ Stress\ =\ Intercept\ +\ Temp.\ +\ Log(\sqrt{Leak-Check\ Pressure}).\ee
	
	Thus, to understand the probability that an O-ring will experience thermal distress at a temperature of $ 31 $ degrees Farenhite, an analysis was done on the O-rings experiencing thermal distress for the 23 shuttle launches prior to the Challenger disaster\footnote{The launch temperature on the day of the Challenger disaster was $ 31 $ degrees Farenhite.}. Thus, I seek to extrapolate the probability of stress, and therefore failure of an O-ring as a function of this temperature. Though the dataset is extremely small, the TrD consisted of the first 18 observations and TeD comprised of the rest.
	
	The results are interesting in that all models compared gave perfect TeD classification. However, the unified methods outperformed the Nonparametric Penalized application in TrD and matched the Nonparametric application in TrD. In regards to AIC, however, the Nonparametric Penalized application had the best TeD result, but the unified methodologies again outperformed the nonunified methods on average over both TrD and TeD combined. Once again, as for the other dataset applications, the main goal here is to compare the unified and nonunified methodologies, and as such the comparisons for other methodologies are not explicitly considered, though the results are consistent with previous findings for these applications as well. Note that the Parametric Logistic is run for only $ 1,000 $ iterations, but the Bayesian Latent Probit is run for $ 5,000 $ iterations, as such, these results are not directly comparable. I refer the reader to \cite{CHOWDHURY2021101112} for more discussions on this. Finally, in regards to inference, each of the unified methodologies were consistent in finding both temperature and pressure as significant, but not the intercepts. In contrast, the Proposed Nonparametric methodology found the intercept as significant as well as temperature and pressure, but the Proposed Penalized Nonparametric methodology found all except the intercept as significant. Accordingly, the results of the unified methodologies are again more consistent than the other models compared. More discussions on this and its implications are given in Section \ref{unifdisc}.
	\begin{figure}[!htb]
		\begin{minipage}{0.5\textwidth}
			\centering
			\includegraphics[width=0.8\linewidth]{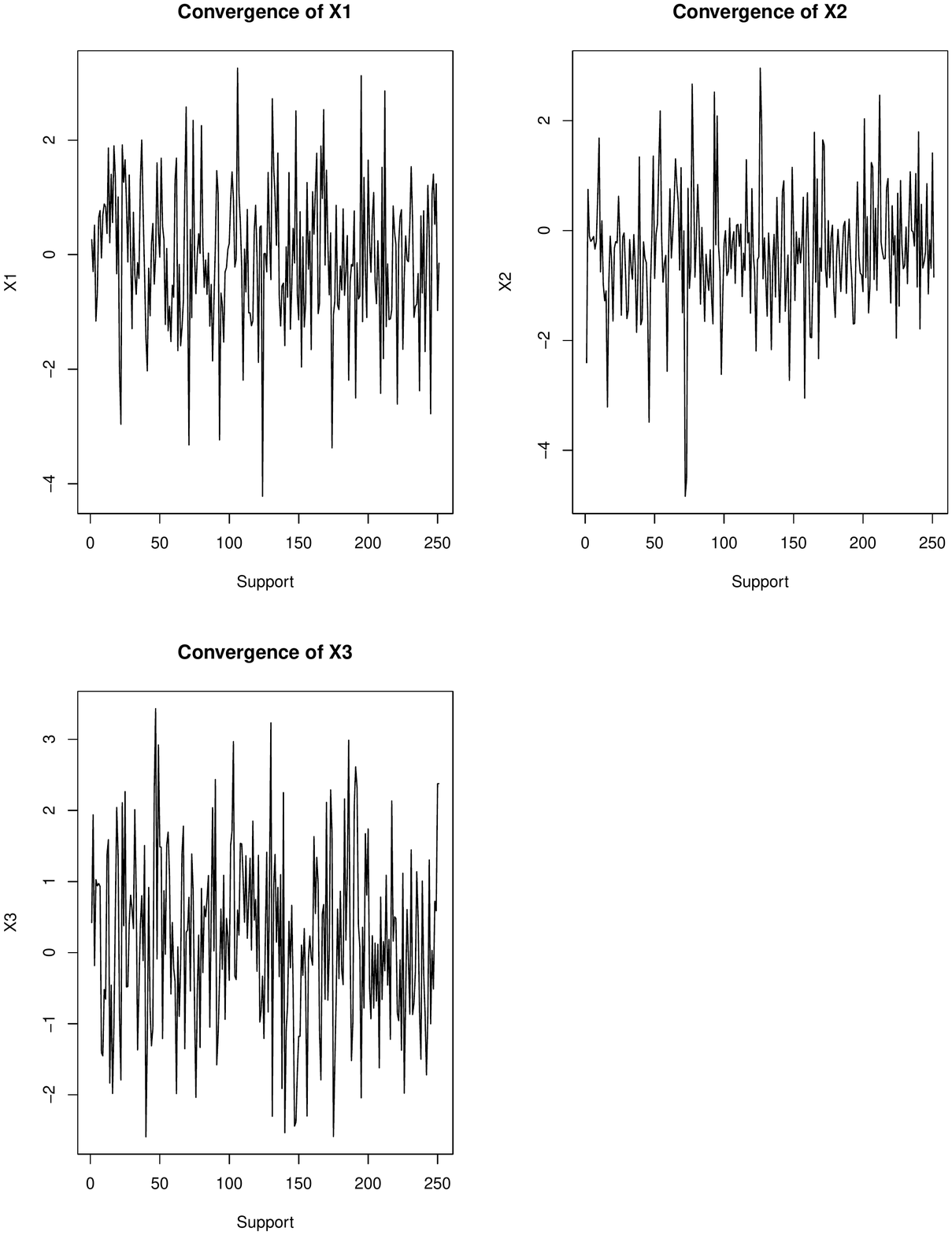}
			\caption{Sample Challenger Sample Space  Exploration Plot for Nonparametric Methodology.}
			\label{fig:chconv1c4}
		\end{minipage}%
		\begin{minipage}{0.5\textwidth}
			\centering
			\includegraphics[width=0.8\linewidth]{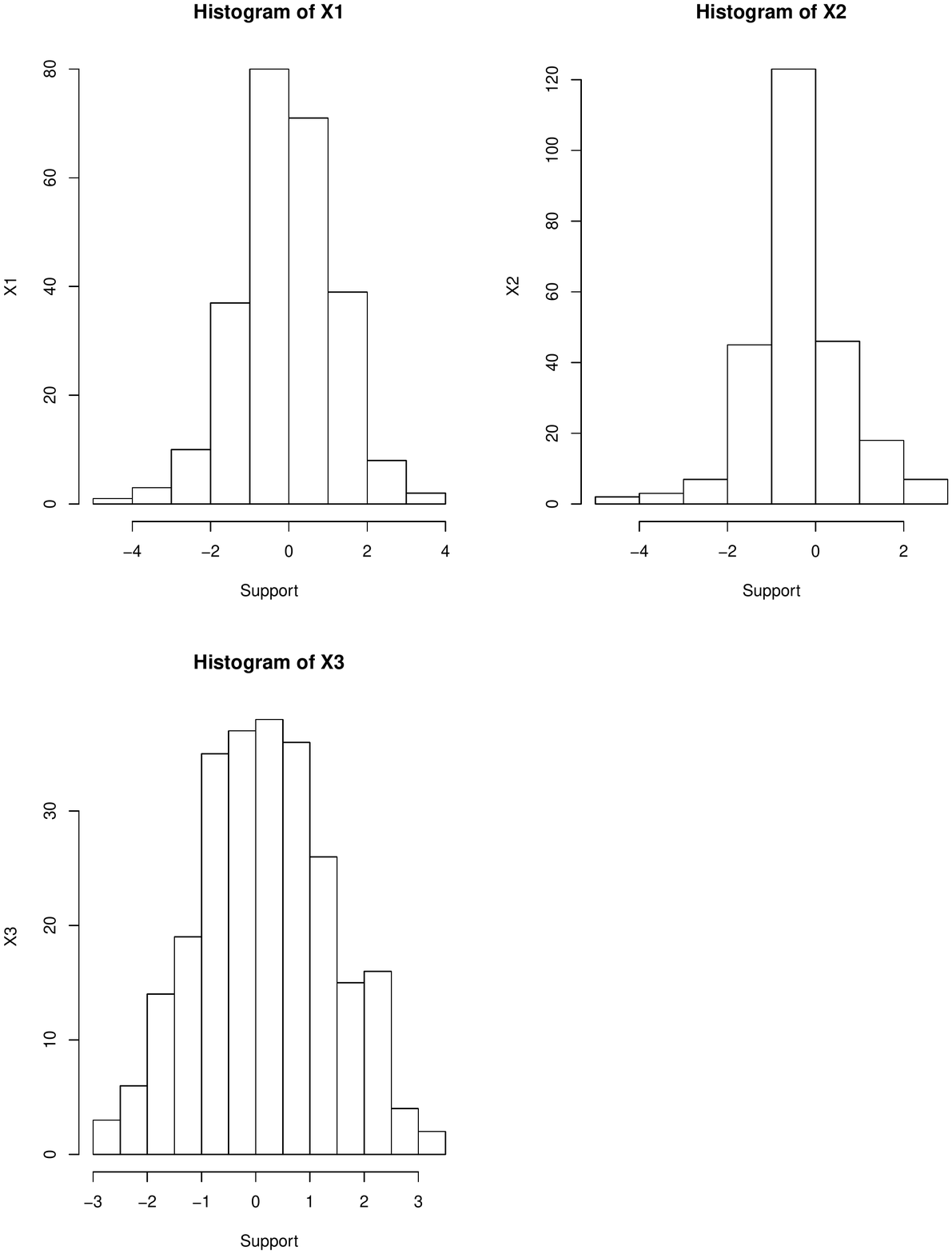}
			\caption{Sample Challenger Histogram of Parameters for Nonparametric Methodology.}
			\label{fig:chhist2c4}
		\end{minipage}
	\end{figure}
	\begin{figure}[!htb]
		\begin{minipage}{0.5\textwidth}
			\centering
			\includegraphics[width=0.8\linewidth]{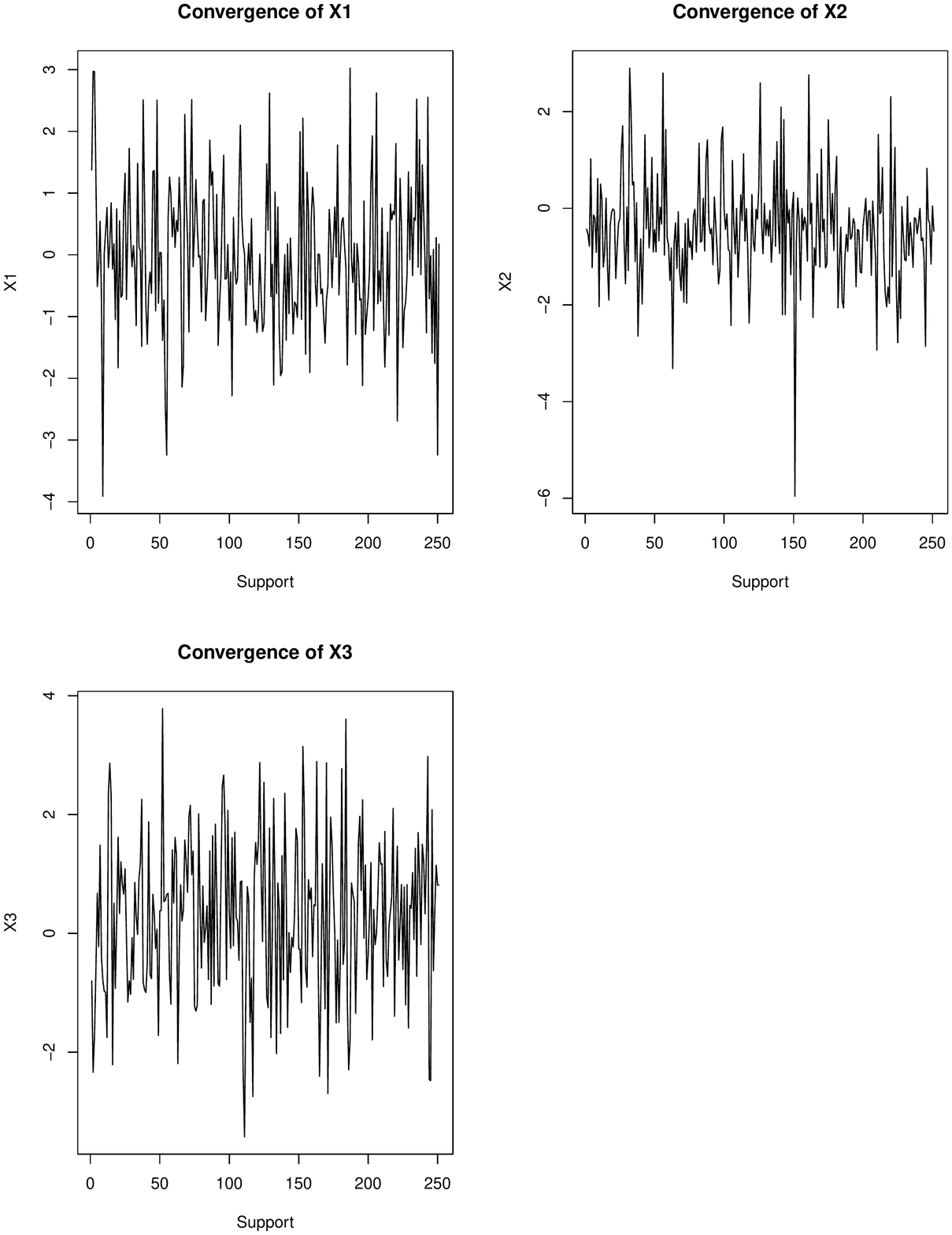}
			\caption{Sample Challenger Sample Space Exploration Plot for Nonparametric Penalized Methodology.}
			\label{fig:chconv3c4}
		\end{minipage}%
		\begin{minipage}{0.5\textwidth}
			\centering
			\includegraphics[width=0.8\linewidth]{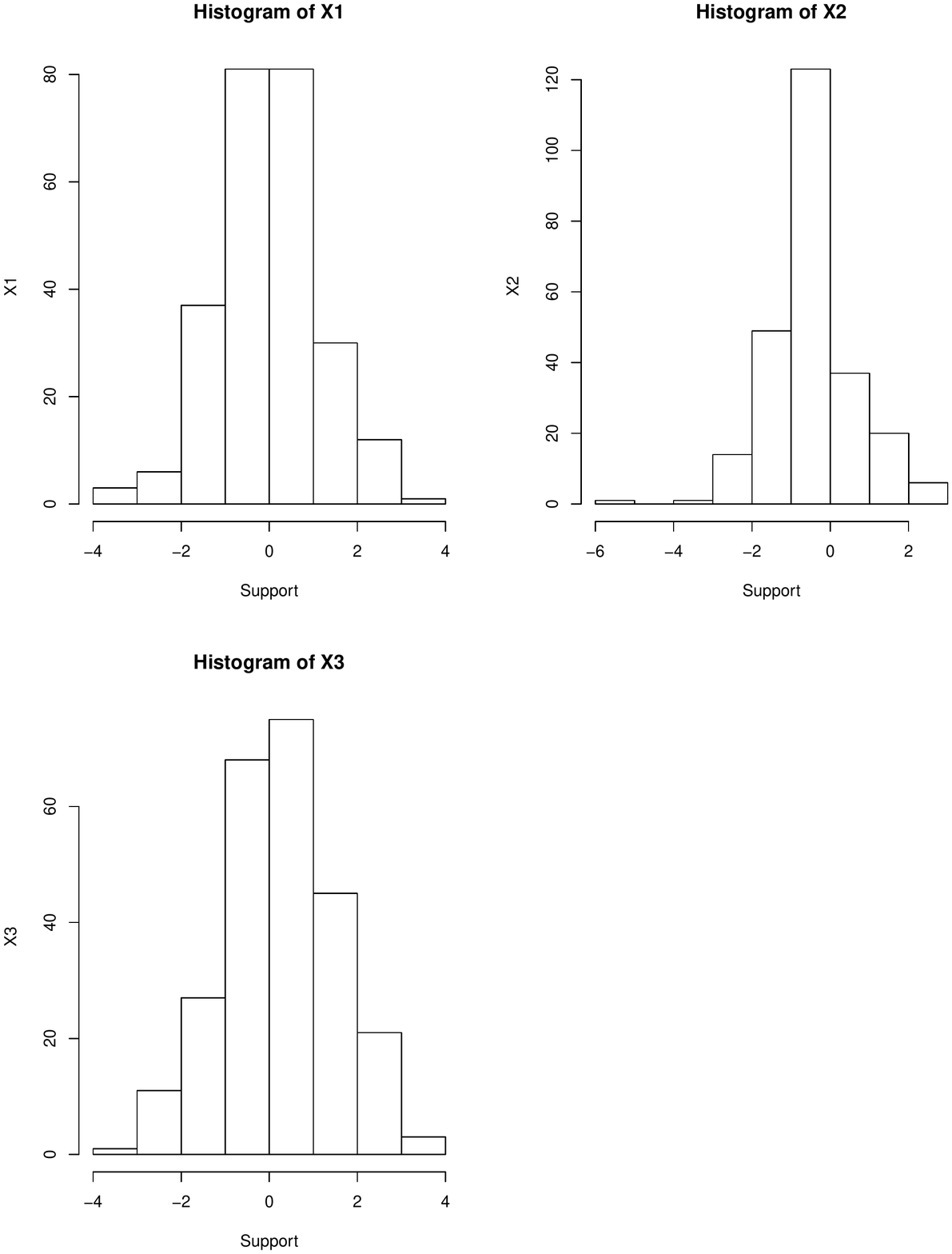}
			\caption{Sample Challenger Histogram of Parameters for Nonparametric Penalized Methodology.}
			\label{fig:chhist4c4}
		\end{minipage}
	\end{figure}
	\begin{figure}[!htb]
		\begin{minipage}{0.5\textwidth}
			\centering
			\includegraphics[width=0.8\linewidth]{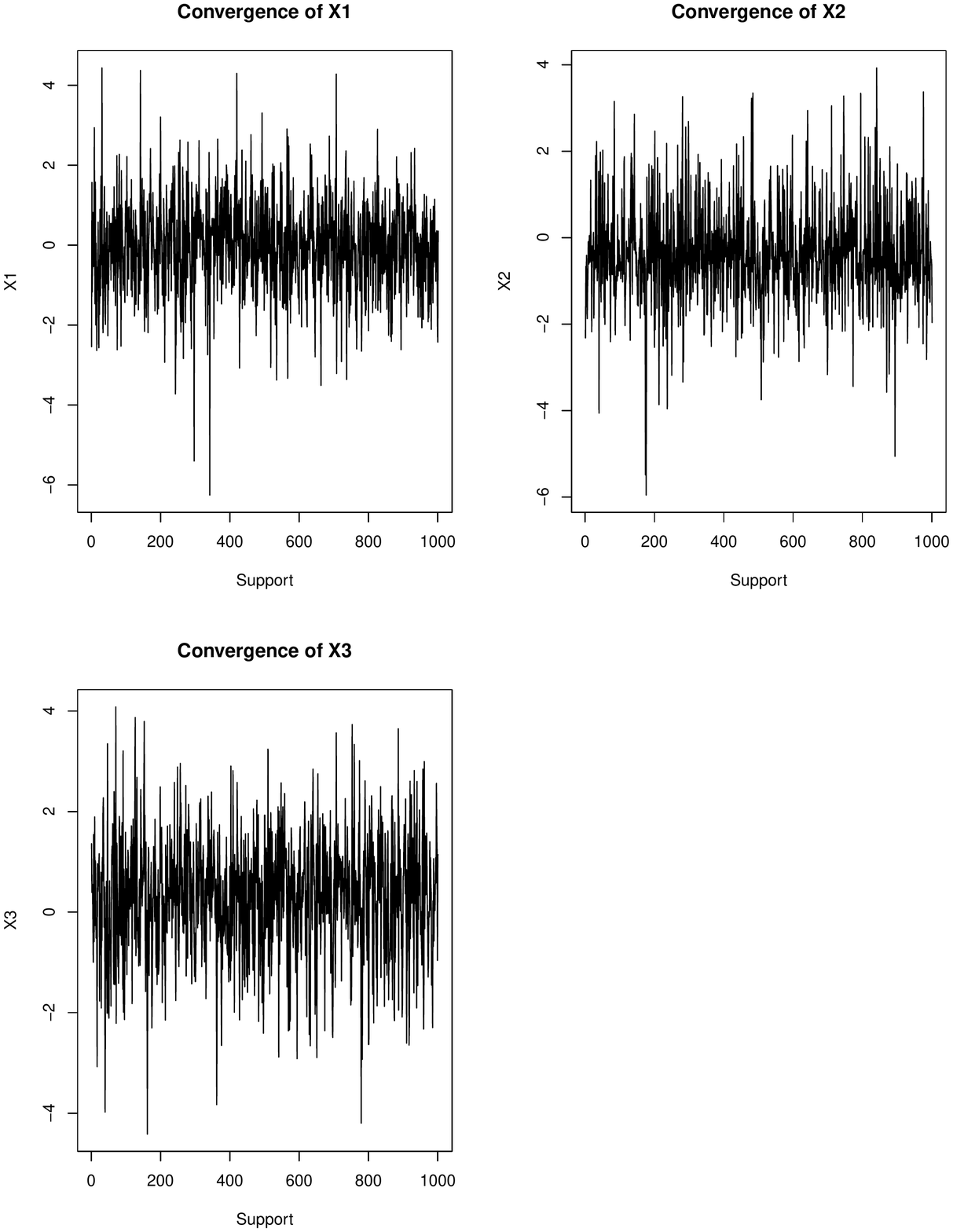}
			\caption{Challenger Sample Space \newline Exploration Plot for Nonparam-\newline etric Unified Methodology.}
			\label{fig:chconv5c4}
		\end{minipage}%
		\begin{minipage}{0.5\textwidth}
			\centering
			\includegraphics[width=0.8\linewidth]{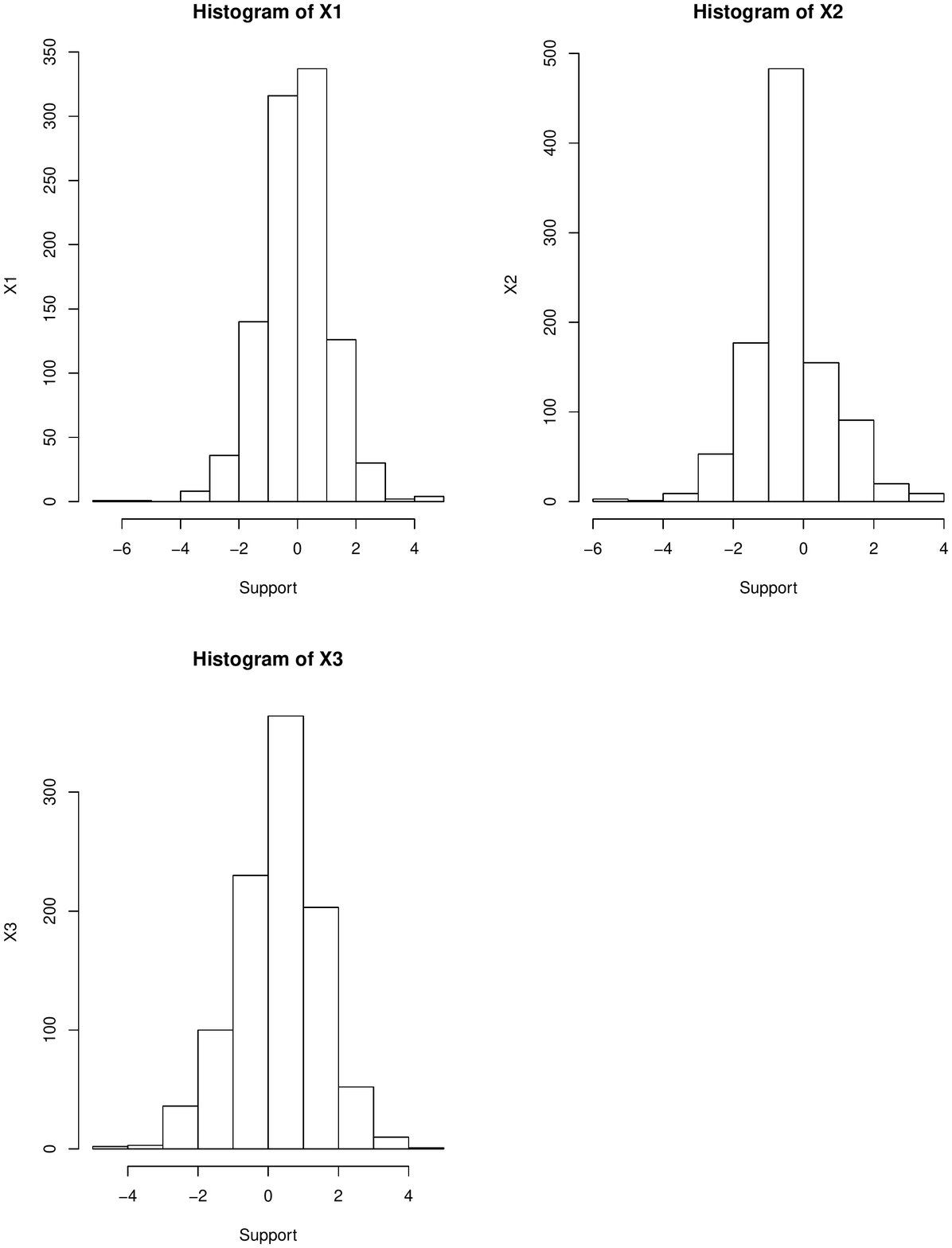}
			\caption{Challenger Histogram of Parameters for Nonparametric Unified Methodology.}
			\label{fig:chhist6c4}
		\end{minipage}
	\end{figure}
	\begin{figure}[!htb]
		\begin{minipage}{0.5\textwidth}
			\centering
			\includegraphics[width=0.8\linewidth]{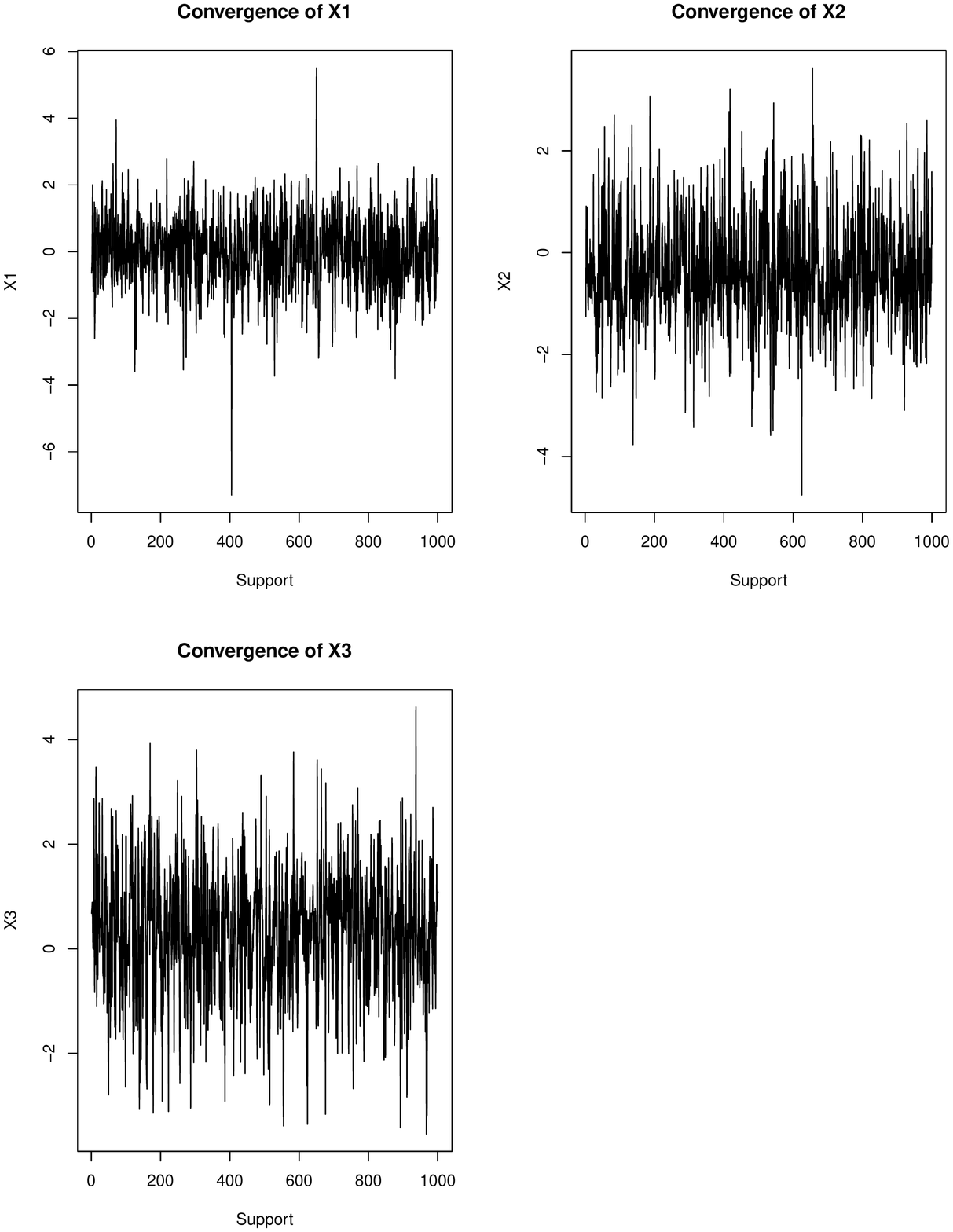}
			\caption{Challenger Sample Space \newline Exploration Plot for Nonparametric \newline Penalized Unified Methodology.}
			\label{fig:chconv7c4}
		\end{minipage}%
		\begin{minipage}{0.5\textwidth}
			\centering
			\includegraphics[width=0.8\linewidth]{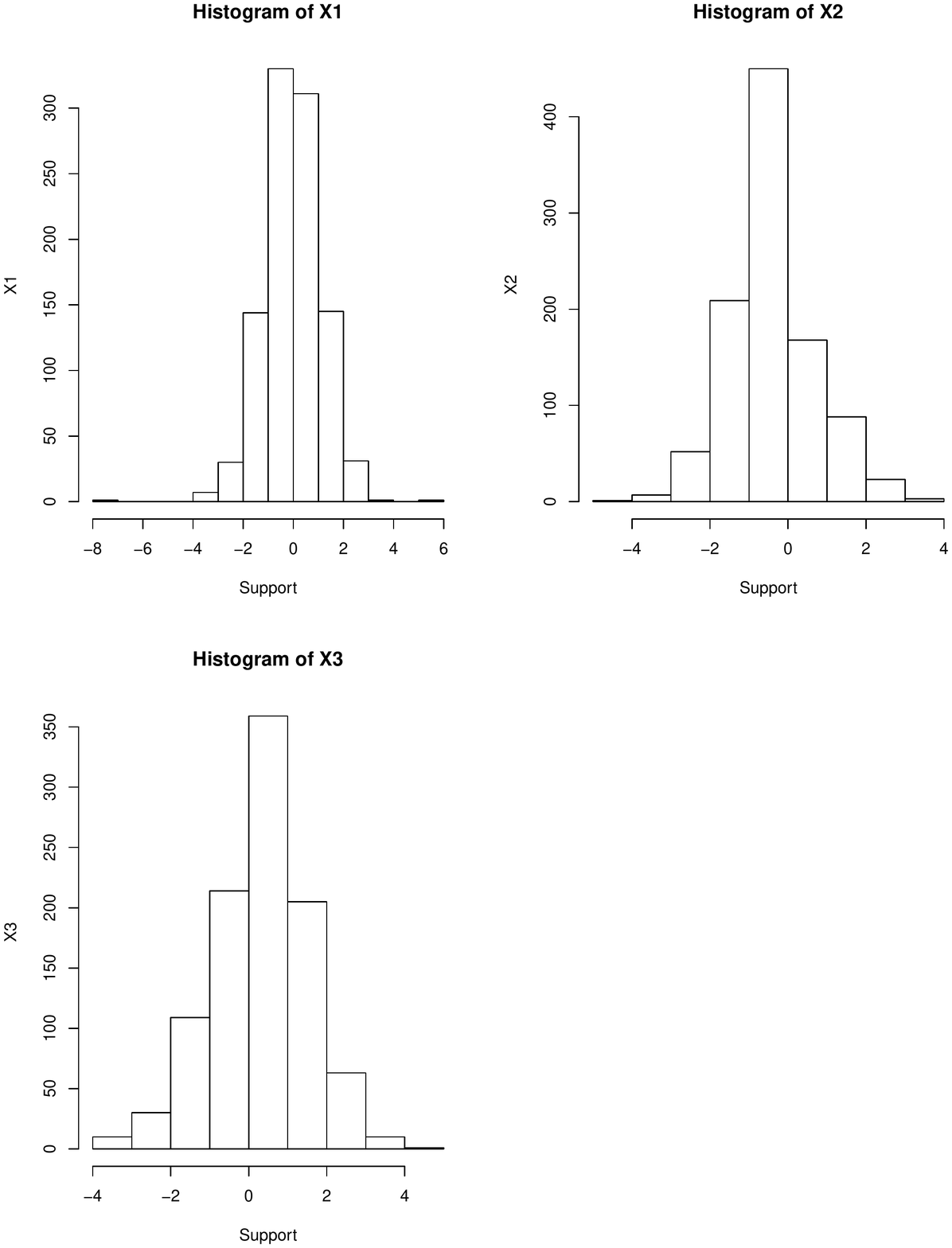}
			\caption{Challenger Histogram of Parameters for Nonparametric Penalized Unified Methodology.}
			\label{fig:chhist8c4}
		\end{minipage}
	\end{figure}
	\begin{table}[h!]
		\begin{minipage}{0.5\textwidth}
			\caption{Challenger Dataset Summary of ARS for All Relevant Methodologies}
			\begin{center}
					\begin{tabular}{ccc}\hline\hline
						Methodology&TrD&TeD\\
						\hline
						Unified Penalized  & 0.00 & 0.00 \\
						Penalized Nonparametric & 0.33 & 0.00 \\
						Nonparametric & 0.00 & 0.00 \\
						Unified Nonparametric & 0.00 & 0.00 \\
						Parametric & 0.34 & 0.00 \\
						Existing Bayes & 0.41 & 0.00 \\
						MLE Logistic & 0.41 & 0.00 \\
						Penalized Logistic & 0.41 & 0.00 \\
						\hline
					\end{tabular}
			\end{center}
			\label{ChallARS}
			\scriptsize{} 
		\end{minipage}\begin{minipage}{0.5\textwidth}
			\caption{Challenger Dataset Summary of AIC for All Relevant Methodologies}
			\begin{center}
					\begin{tabular}{ccc}\hline\hline
						Methodology&TrD&TeD\\
						\hline
						Unified Penalized & 0.96 & 1.25 \\
						Penalized Nonparametric& 3.16 & 0.95 \\
						Nonparametric & 2.20 & 1.42 \\
						Unified Nonparametric& 1.02 & 1.25 \\
						Parametric & 1.43 & 2.41 \\
						Existing Bayes & 2.63 & 2.57 \\
						MLE Logistic & 1.09 & 2.06 \\
						Penalized Logistic & 1.24 & 2.50 \\
						\hline
					\end{tabular}
			\end{center}
			\label{ChallAIC}
			\scriptsize{} 
		\end{minipage}
	\end{table}
	\begin{table}[!h]
		\begin{minipage}{1\textwidth}
			\caption{Challenger Dataset Parameter Summary for All Relevant Methodologies}
			\begin{center}
					\begin{tabular}{ccccc}\hline\hline
						Predictor&Estimates&CI-Low&CI-High&Methodology\\
						\hline
						Intercept & 0.11 & -0.3 & 0.52 & \multirow{3}{*}{(1)} \\
						Temperature & -0.4** & -0.73 & -0.08 &\\
						Pressure & -0.49** & -0.81 & -0.18 & \\
						Intercept & -0.06** & -0.12 & -0.01 &  \multirow{3}{*}{(2)}\\
						Temperature & -0.54** & -0.59 & -0.5 & \\
						Pressure & 0.34** & 0.28 & 0.40 & \\
						Intercept & -0.15 & -0.48 & 0.18 & \multirow{3}{*}{(3)}\\ 
						Temperature & -0.55** & -0.86 & -0.24 &\\ 
						Pressure & 0.42** & 0.05 & 0.79 & \\
						Intercept & 0.06 & -0.09 & 0.21 & \multirow{3}{*}{(4)}\\
						Temperature & -0.45**& -0.60 & -0.31 & \\
						Pressure & 0.48** & 0.3 & 0.65 & \\
						Intercept & -0.27** & -0.42 & -0.12 & \multirow{3}{*}{(5)}\\
						Temperature & -0.45** & -0.62 & -0.28 &\\
						Pressure & -0.35** & -0.49 & -0.21 & \\
						Intercept & -1.52** & -1.64 & -1.4 & \multirow{3}{*}{(6)}\\
						Temperature & -0.63** & -0.66 & -0.59 & \\
						Pressure & 0.67** & 0.60 & 0.74 & \\
						Intercept & -4.5 & -14.57 & 5.56 & \multirow{3}{*}{(7)}\\
						Temperature & -1.97* & -4.31 & 0.38 &\\
						Pressure & 1.35 & -2.86 & 5.56 & \\
						Intercept & -3.65 & -17.89 & 7.2 & \multirow{3}{*}{(8)}\\
						Temperature & -1.24** & -5.09 & 0.2 &\\
						Pressure & 1.11 & -3.71 & 6.76 & \\
						\hline
					\end{tabular}
			\end{center}
			\label{ChallParam}
			\scriptsize{Note: (1) Nonprametric, (2) Unified Nonparametric, (3) Penalized Nonparametric, (4) Unified Penalized Nonparametric, (5) Parametric, (6) Existing Bayesian (7) MLE Logistic (8) Penalized Logistic.} 
		\end{minipage}
	\end{table}
	\clearpage
	\section{Unique Random Utility Elicitation}
	
	In this section I highlight a social science application. The field is vast on this subject and I focus the discussion specifically on Additive Random Utility Models (ARUM). The premise of the formulation is both useful and intuitive, and as such is used widely across the social sciences. For completeness I give a brief introduction below. Let $ y $ take the values of $ 0 $ or $ 1 $ between two alternatives, where if an agent picks an alternative it is represented by $ y $ being 1. Then following \cite{cameron2010microeconometrics} we may write the utilities gained from each alternative, measured with some error as, \begin{eqnarray}
		U_0 = V_0 +\epsilon_0,\\
		U_1 = V_1 + \epsilon_1.
	\end{eqnarray}
	The only random component here are $ \{\epsilon_0, \epsilon_1\} $. Thus, a person chooses alternative $ 1 $ over alternative $ 0 $ if, \begin{eqnarray}
		Pr[y=1] = Pr[U_1 > U_0] = F(\epsilon_0 - \epsilon_1 < V_1 - V_0) = F(\b x\b\beta). 
	\end{eqnarray}
	Considering a simple linear model such that, \e
	V_1= \b x\b\beta_1; V_0 = \b x\b \beta_0 \implies V_1 - V_0 = \b x\b\beta.
	\ee
	Our formulation provides us with a decided advantage over the existing formulation in this framework because, we may look at translation invariant measures. Which in turn ensures that we do not need to hold the variance constant, while getting strongly convergent parameters. That is, since utility is an ordinal concept, and since we have an unique translation invariant measure as a function of the strongly convergent parameters $ \b \beta $, we may consider $ V_0 = \b x\b \beta_0 = 0 $, such that \e
	V_1 = \b x \b \beta.
	\ee 
	But, $ \b \beta $ is strongly convergent and unique in our formulation. Therefore, $ V_1 $ under our preliminaries is unique such that, \e
	U_1 = E_{\b \beta}(V_1|\b x) = \b x \b \beta,
	\ee
	where, as before we did not need to fix our variance in this formulation. Of course, the formulation can easily be applied in another functional formulation such as $ c(X)\beta $ as well, with the variance being a function of the functional specification. Therefore, the most general formulation has given us unique first and second moments of the unique distribution without any loss of generality! The preliminaries are also entirely consistent with utility theory, such that setting the baseline alternative utility to a specified number does not result in any loss of information. Therefore, though utility is an ordinal concept, if we fix the baseline category utility to a specific number, all other utilities may be compared cardinally with respect to this baseline utility. Therefore, that we set the baseline utility in our formulation to $ 0 $, is in effect without loss of any particular generality. As such $ U_1 $ is unique as well!
	
	To apply this in a simple formulation I consider a Bank Marketing Dataset accessed from the UCI Machine Learning Repository, \href{https://archive.ics.uci.edu/ml/datasets/Bank+Marketing}{Center for Machine Learning and Intelligent Systems} (last accessed 12-14-2021). The dataset consists of outcomes from marketing campaigns for a Portugese banking institution, with the goal to get them to subscribe to term deposits at the bank as in \cite{moro2014data}. There are a number of variables and for greater details the interested reader is referred to the article. For our present purpose we would like to simply understand how effective our simple model is in regards to MIP. Accordingly, consider the model, \vspace{-.2in}
	\begin{center}
		\begin{align}
			Deposit = &Intercept + Job + Marital\ Status + Education\ Level + \nonumber \\
			&Default\ Status + Balance + Housing\ Loan \nonumber \\
			&+ Loan\ Status + Campaign.
		\end{align}
	\end{center}
	\begin{table}[h!]
		\begin{minipage}{0.5\textwidth}
			\caption{Banking Dataset Summary of ARS for All Relevant Methodologies}
			\begin{center}
					\begin{tabular}{ccc}\hline\hline
						Methodology&TrD&TeD\\
						\hline
						Unified Penalized  & 0.27 & 0.05 \\
						Penalized Nonparametric & 0.37 & 0.00 \\
						Nonparametric & 0.45 & 0.03 \\
						Unified Nonparametric & 0.36 & 0.06 \\
						Parametric & 0.67 & 0.70 \\
						Existing Bayes & 0.72 & 0.79 \\
						MLE Logistic & 0.62  & NA \\
						Penalized Logistic & 0.62 & NA \\
						\hline
					\end{tabular}
			\end{center}
			\label{UUARS}
			\scriptsize{} 
		\end{minipage}\begin{minipage}{0.5\textwidth}
			\caption{Banking Dataset Summary of AIC for All Relevant Methodologies}
			\begin{center}
					\begin{tabular}{ccc}\hline\hline
						Methodology&TrD&TeD\\
						\hline
						Unified Penalized & 1.40 & 1.27 \\
						Penalized Nonparametric& 3.77 & 1.39 \\
						Nonparametric & 4.01 & 5.11 \\
						Unified Nonparametric& 1.62 & 1.30 \\
						Parametric & 1.23 & 1.10 \\
						Existing Bayes & 4.00  & 3.42 \\
						MLE Logistic & 2.42 & NA  \\
						Penalized Logistic & 2.42 & NA  \\
						\hline
					\end{tabular}
			\end{center}
			\label{UUAIC}
			\scriptsize{} 
		\end{minipage}
	\end{table}
	\begin{figure}
		\begin{minipage}{.5\textwidth}
			\centering
			\includegraphics[width=0.7\linewidth]{"BankConv"}
			\caption{Sample Space Exploration \newline Bank Data}
			\label{fig:bankconv}
		\end{minipage}\begin{minipage}{.5\textwidth}
			\centering
			\includegraphics[width=0.7\linewidth]{"BankHist"}
			\caption{Histogram of Parameters of \newline Bank Data}
			\label{fig:bankhist}
		\end{minipage}
	\end{figure}
	\begin{figure}
		\begin{minipage}{.5\textwidth}
			\centering
			\includegraphics[width=0.7\linewidth]{"BankPenConv"}
			\caption{Penalized Sample Space Exploration Bank Data}
			\label{fig:bankpenconv}
		\end{minipage}\begin{minipage}{.5\textwidth}
			\centering
			\includegraphics[width=0.7\linewidth]{"BankPenHist"}
			\caption{Penalized Histogram of Parameters of Bank Data}
			\label{fig:bankpenhist}
		\end{minipage}
	\end{figure}
	\begin{figure}
		\begin{minipage}{.5\textwidth}
			\centering
			\includegraphics[width=0.7\linewidth]{"BankPenUnifConv"}
			\caption{Unified Penalized Sample Space Exploration Bank Data}
			\label{fig:bankpenunifconv}
		\end{minipage}\begin{minipage}{.5\textwidth}
			\centering
			\includegraphics[width=0.7\linewidth]{"BankPenUnifHist"}
			\caption{Unified Penalized Histogram of Parameters of Bank Data}
			\label{fig:bankpenunifhist}
		\end{minipage}
	\end{figure}
	\begin{figure}
		\begin{minipage}{.5\textwidth}
			\centering
			\includegraphics[width=0.7\linewidth]{"BankUnifConv"}
			\caption{Unified Sample Space Exploration Bank Data}
			\label{fig:bankunifconv}
		\end{minipage}\begin{minipage}{.5\textwidth}
			\centering
			\includegraphics[width=0.7\linewidth]{"BankUnifHist"}
			\caption{Unified Histogram of Parameters of Bank Data}
			\label{fig:bankunifhist}
		\end{minipage}
	\end{figure}

\begin{figure}
		\begin{minipage}{1\textwidth}
			\centering
			\includegraphics[width=0.25\linewidth]{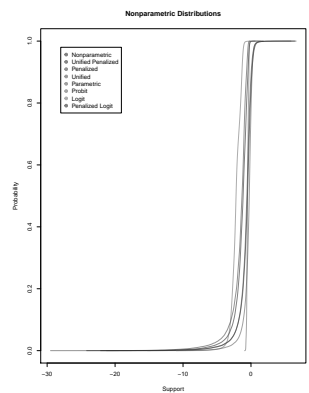}
			\caption{Nonparametric Distributions Based on Converged Parameters by Method}
			\label{fig:distributions}
		\end{minipage}
	\end{figure}

The nonparametric distributions based on the converged parameters for each method can be found in Figure \ref{fig:distributions}.
	
	\begin{table}[!h]
	\begin{minipage}{1\textwidth}
					\caption{Bank Dataset Parameter Summary for Relevant Methodologies (I)}
		\begin{center}
				\begin{tabular}{ccccc}\hline\hline
					Parameters&Estimate&CI-Low&CI-High&Methodology\\
					\hline
					Intercept & 0.07 & -0.04 & 0.17 &\multirow{9}{*}{(1)}  \\
					Job& -0.13** & -0.23 & -0.03 & \\
					Marital Status & -0.01 & -0.11 & 0.10 &  \\
					Education Level & 0.06 & -0.04 & 0.16 &  \\
					Default Status & -0.04 & -0.14 & 0.06 & \\
					Balance & 0.06 & -0.04 & 0.16 &  \\
					Housing Loan & 0.00 & -0.1 & 0.11 & \\
					Loan Status & -0.09 & -0.19 & 0.01 & \\
					Campaign & -0.39** & -0.49 & -0.28 &  \\
					\hline
					Intercept & -0.05** & -0.07 & -0.03 & \multirow{9}{*}{(2)} \\
					Job & 0.08** & 0.06 & 0.11 &  \\
					Marital Status & -0.03** & -0.05 & -0.01 &  \\
					Education Level & -0.19** & -0.21 & -0.16 &  \\
					Default Status & 0.01 & -0.01 & 0.03 & \\
					Balance & 0.19** & 0.17 & 0.22 & \\
					Housing Loan & 0.18** & 0.16 & 0.20 &  \\
					Loan Status & -0.03** & -0.05 & -0.01 &  \\
					Campaign & -0.34** & -0.36 & -0.31 & \\
					\hline
					Intercept & -0.03 & -0.13 & 0.07 & \multirow{9}{*}{(3)} \\
					Job & -0.08 & -0.18 & 0.02 &  \\
					Marital Status & -0.09 & -0.19 & 0.02 &  \\
					Education Level & -0.09 & -0.19 & 0.02 & \\
					Default Status & 0.00 & -0.1 & 0.11 &  \\
					Balance & -0.01 & -0.1 & 0.09 &  \\
					Housing Loan & 0.06 & -0.05 & 0.16 &  \\
					Loan Status & 0.06 & -0.04 & 0.17 &  \\
					Campaign & -0.46** & -0.57 & -0.35 &  \\
					\hline
					Intercept & 0.07 & -0.03 & 0.18 & \multirow{9}{*}{(4)} \\
					Job & 0.07 & -0.03 & 0.17 &  \\
					Marital Status & 0.12** & 0.01 & 0.23 &  \\
					Education Level & -0.18** & -0.28 & -0.09 &  \\
					Default Status & 0.01 & -0.09 & 0.11 &  \\
					Balance & 0.20** & 0.09 & 0.30 &  \\
					Housing Loan & 0.23** & 0.13 & 0.34 &  \\
					Loan Status & -0.15** & -0.25 & -0.05 &  \\
					Campaign & -0.35** & -0.45 & -0.26 &  \\
					\hline
					\hline
				\end{tabular}
		\end{center}
		\label{IntxParamI}
		\scriptsize{Note: (1) Nonprametric, (2) Unified Nonparametric, (3) Penalized Nonparametric, (4) Unified Penalized Nonparametric. ** indicates significant at the 0.01 level.} 
	\end{minipage}
\end{table}
\clearpage

\begin{table}[h!]
\begin{minipage}{1\textwidth}
		\caption{Bank Dataset Parameter Summary for Relevant Methodologies (II)}
		\begin{center}
				\begin{tabular}{ccccc}\hline\hline
					Parameters&Estimate&CI-Low&CI-High&Methodology\\
					\hline
					Intercept & 0.15** & 0.05 & 0.24 & \multirow{9}{*}{(5)} \\
					Job & 0.03 & -0.04 & 0.1 &  \\
					Marital Status & 0.03 & -0.07 & 0.13 &  \\
					Education Level & -0.07 & -0.17 & 0.02 & \\
					Default Status & -0.03 & -0.12 & 0.07 &  \\
					Balance & 0.17** & 0.08 & 0.26 &  \\
					Housing Loan & 0.23** & 0.13 & 0.33 &  \\
					Loan Status & -0.12** & -0.22 & -0.02 &  \\
					Campaign & -0.25** & -0.33 & -0.16 &  \\
					\hline
					Intercept & -0.47** & -0.52 & -0.41 & \multirow{9}{*}{(6)} \\
					Job & -0.02** & -0.04 & 0.00 &  \\
					Marital Status & 0.00 & -0.04 & 0.04 &  \\
					Education Level & -0.03 & -0.06 & 0.00 & \\
					Default Status & -0.01 & -0.12 & 0.10 & \\
					Balance & 0.03 & -0.01 & 0.06 &  \\
					Housing Loan & 0.18** & 0.14 & 0.23 &  \\
					Loan Status & -0.14** & -0.19 & -0.08 &  \\
					Campaign & 0.00 & -0.01 & 0.02 &  \\
					\hline
					Intercept & -1.96** & -2.04 & -1.88 &\multirow{9}{*}{(7)}\\
					Job & 0.02** & 0.01 & 0.03 &  \\
					Marital Status & 0.08** & 0.03 & 0.12 & \\
					Education Level & -0.19** & -0.23 & -0.16 & \\
					Default Status & -0.55** & -0.88 & -0.22 &\\
					Balance & 0.08** & 0.05 & 0.1 & \\
					Housing Loan & 0.85** & 0.78 & 0.92 & \\
					Loan Status & -0.63** & -0.74 & -0.52 & \\
					Campaign & -0.14** & -0.16 & -0.12 & \\
					\hline
					Intercept & -1.96** & -2.07 & -1.86 & \multirow{9}{*}{(8)}\\
					Job & 0.02** & 0.01 & 0.04 & \\
					Marital Status & 0.08** & 0.01 & 0.14 & \\
					Education Level & -0.19** & -0.24 & -0.14 &\\
					Default Status & -0.54** & -1.00 & -0.12 & \\
					Balance & 0.08** & 0.04 & 0.11 & \\
					Housing Loan & 0.85** & 0.76 & 0.94 &  \\
					Loan Status & -0.63** & -0.78 & -0.49 & \\
					Campaign & -0.14** & -0.16 & -0.11 & \\				
					\hline
					\hline
				\end{tabular}
		\end{center}
		\label{IntxParamII}
		\scriptsize{Note: (5) Parametric, (6) Existing Bayesian, (7) MLE Logistic, (8) Penalized Logistic. ** indicates significant at the 0.01 level.} 
	\end{minipage}
\end{table}

	\vspace{.2in}
	
	The results are interesting and may at first seem contradictory since the significant variables change both sign and significance from one model to another. However, this is another example of the importance of a wholistic approach to MIPs. For example, it seems rather implausible that all the variables would be so significant as in the Logistic $ (7) $ and  Penalized Logistic $ (8) $ cases. On the other hand the proposed methods also seem to give different signs for each of the variables. Yet if we look at all the MIP criterias, there is a clear winner here and that is the Unified Penalized Nonparametric methodology. To see this note that it has the lowest average ARS, which is significantly lower than the Parametric version, with only slightly worse model fit than the Parametric methodology. Thus, if we were to look at the significance results for the Unified Penalized version, we see that it finds Marital Status, Education Level, Balance, Housing Loan, Loan Status, and Campaign as significant. These results seem to be consistent based on the numerical set up of the variables themselves, since all but the Balance variable are categorical. What is consistent across the models is that higher number of marketing contacts (Campaign) seems to have reduced Deposit outcomes. An interesting results indeed. Thus, for managers instead of focusing on increased marketing contacts in this case they should instead focus on more meaningful contacts that outlines the value propositions clearly to the would be clients. Remarkebly, while the original paper found neural networks to have the best predictive outcomes with a AUC of 0.8, all of the proposed methods outperformed that level of AUC performance (Figure \ref{UUAUC}). Even without using all the available variables, in a very simple functional formulation. Thus, it is very plausible that near perfect/perfect AUC outcomes could have been achieved here. However, this is not further pursued here.
	\begin{table}[h!]
		\begin{minipage}{1\textwidth}
			\caption{Banking Dataset Summary of AUC for Proposed Methodologies}
			\begin{center}
					\begin{tabular}{ccc}\hline\hline
						Methodology&TrD&TeD\\
						\hline
						Unified Penalized  & 0.93 & 0.94 \\
						Penalized Nonparametric & 0.80 & 0.97 \\
						Nonparametric & 0.82 & 0.97 \\
						Unified Nonparametric & 0.90 & 0.94 \\
						\hline
					\end{tabular}
			\end{center}
			\label{UUAUC}
			\scriptsize{} 
		\end{minipage}
	\end{table}

	\section{Discussion}\label{unifdisc}
	
The results above are noteworthy in several contexts, including for Model Fit, Inference, and Prediction. Thus in the forthwith I will expand on the importance of these findings for our understanding of how statistical significance is related to scientific significance. However, before discussing these topics I would first like to highlight some of the mathematical results that allow us to make these conclusions that have broad impact across the sciences.

One of the most important results of this formulation is that the formulation ensures that mathematically a linear operator exists between the Hausdorff space and the link field. The claim is significant in that as sample size increases, in the current framework we cannot guarantee that the estimation process, whether Bayesian or Frequentist, will be bounded for all observations. Thus, though this occurs at the unbounded points in the Hausdorff space, we cannot assert that this occurrence will have measure 0. However, the formulation presented here ensures, under measure theoretic foundations, using novel analytic results, that the estimation process will be continuous. Moreover, it assures that such a linear operator will strongly converge to the true parameters. LAHEML therefore ensures that as a function of the model specification and the observed Xs, convergence of the parameters of interest will be almost sure. This is an expansive result which many existing convergence methodologies cannot claim, because in their estimation the assertion may be violated. In addition, the construction by definition ensures that the variance parameter may easily be computed accordingly. As such, this has some rather strong implications for our continuing discussion on statistical significance and its relation to scienctific significance, and I will expand on this more below.

Another surprising result is that though the formulation may seem categorical in nature, the results of \cite{nonparafuncarxiv} ensure that as the number of observations increase it can also be applicable to continuous outcome models. The mathematical results of \cite{stronglyconvergentMestarxiv} reinforce those results.  Thus, as almost sure convergence results are for general functional forms, they are also valid and applicable for continuous outcome data. As such, the methodologies presented above overcome one of the principle shortcomings of mathematical model estimations present in the sciences for many decades. 

Indeed, in many ways the formulation is not intuitive at first glance. Since, it would seem counterintuitive to use a signed measure to estimate functional forms over $ \sigma-algebras $ that are a function of probability distributions. Yet the locally compact Hausdorff formulation ensures the most general topological circumstances under which such a construction remains valid. In fact, it is easy to see that in existing latent variable formulations, in the absence of continuity of the linear operator, the conditions for convergence do not hold in \cite{tanner1987calculation}. This is because, in such a case the underlying functional specifications are not equicontinuous. Thus, it is nested within the present formulation and its implementation through LAHEML.  

Evidently, the use of a measure theoretic approach rooted in functional and real analysis, thus have demonstrable advantages over the existing formulations. This is because, through it, we may consider the $ L^p(X,\nu) $ spaces for $ 1 \le p \le \infty, $ where as before $ p= \infty $ is the essentially bounded case. Using the locally compact Hausdorff property we are able to show that there exists a linear operator on the $ L^p $ spaces, which are also Banach spaces which are an isometric isomorphism to the space of bounded finitely additive signed measures. As such, we are able to apply the methodology on stronger Banach Spaces such as $ L^{p=1} $ which contains the set of functions in the other $ L^p $ spaces. Consequently, it readily allows more flexibility in the potential functions that we may use to optimize over. On the other hand, positive finiteness almost everywhere, in conjunction with a locally compact Hausdorff property, implies that we may identify an unique signed measure using Riesz-Moarkov for every linear functional. Thus, LAHEML uses these existence and uniqueness properties to identify the unique linear operator that is a function of this unique signed-measure, thus explaining the excellent results above. Accordingly, above I extended Riesz-Markov Theorem to the general $ L^p $ spaces for all $ 1 \le p \le \infty $, a result which appears to be new in this formulation in analysis, and it certainly is unique to Mathematical Statistics in a latent variable formulation especially in the Bayesian formulation.

Furthermore, this formulation has another remarkable property which goes beyond the independent and identically distributed model formulations which are the cornerstone of the sciences. To see this first note that through the use of \cite{kass1989approximate} we know that observations in LAHEML can be thought of as conditionally independent. Further note that in the unified formulation we may argue that the sample space may be separated into countable unions of disjoint sets, over which we may define a subspace, such that a bounded finitely additive measure space exists over it. But a finitely additive measure may easily be converted to a distribution, and thus, for each disjoint set we may and do define a separate distribution. Consequently, these distributions need not be identical at all! Whether or not they are will depend on the extension of measure spaces through Hahn-Banach on the entire $ \sigma-algebra. $ Yet the proofs of the existence, uniqueness, and almost sure convergence results are not dependent on any particular distributional assumptions, identical or otherwise! Thus, this the most general of formulations which need not require our data to be either independent or identical!

Of course, as with any model specification, our characterization of the underlying phenomenon is a function of the observed Xs. In addition, there may be circumstances under which the assumptions in \cite{kass1989approximate} may not hold either. In such a case, there are a multitude of other methodologies which can be used in conjunction with the proposed methodologies accordingly. Regardless, the usefulness of the methodology and its general construction shows much potential for broad applicability across the science, including for MIPs. 

Hopefully, it is clear that one of the chief contributions of this research is the realization that just because a model outperforms another in regards one of the MIPs, it does not imply that it will outperform it in another. Indeed, there are numerous examples above where the opposite is true. What then can we glean from these findings, in light of the discussion on the virtues of the methodologies above? Firstly, we saw that it need not be the case that we perform model fit and model selection separately in all cases. For example, we saw in the Intoxication dataset that the Unified Penalized methodology had the best AIC of all the methods compared. In the Challenger dataset it also had the best overall model fit. Furthermore, this level of performance was achieved without any noticeable drop in prediction performance, since it was close to the best methods in this regard. In regards to inference as well, the results for the Unified Penalized methodology remained consistent especially for the Challenger and Bank datasets, where it found both temperature and pressure to be significant for the first and numerous scientifically relevant variables for the Bank dataset.

In considering the other methodologies, we can also see that the unified versions uniformly outperformed the nonunified versions, especially in Prediction, and did so again without sacrificing interpretability of the parameter estimates. The mathematical results above provide solid foundations for these results. However, broadly we may think of the nonunified versions as a specific versions of the unified methodology. Thus, the methodology is able to identify the correct parameters, whether the underlying DGP is symmetric or asymmetric, and in either case does not impose the link function approaching 1 or 0 at the same rate. Furthermore, it has many of the virtues of the nonunified version of \cite{nonparafuncarxiv}, such as not needing to hold the variance constant and having continuous errors as well. Many of the same large-sample tests discussed there can also be applied to this formulation. However, now we may choose to apply the tests separately over the Hahn decomposition sets or together to see which methodology gives the best desired result.

In fact, the findings regarding significance of the nonparametric methodology is also relevant here and demand some further discussion. That is, the ability to perform MIPs using stronger topological spaces imply that we no longer need to sacrifice one mathematical goal for another. This is because a larger space of functions allow us a broader range of possible candidates against which we may optimize our parameters of interest as a function of the Xs. In so doing a scientifically interpretable linear operator may outperform complex learning algorithms without sacrifing interpretability of parameter estimates. Since the parameters converge almost surely as a function of the Xs and the model specification, significance then no longer needs to be a 0-1 answer (ironically). Thus, we may use stronger convergence properties, on stronger topologies to sequentially make a model specification more complex as needed. If, on the other hand inference is not an immediate goal, we may use these same robust properties of the methodologies with existing excellent AI and ML methodologies to give equivalent or better results.

One of the principle contributions of this research is in the insights it provides on the interplay of statistical significance and scientific significance. Focusing solely on large sample results with finite sample sizes can lead to bias and inconsistency of the parameter estimates in general. However, the results highlight that even in large samples the convergence strength of the methodology is crucial to scientfically rely on statistical inferential results in a robust way. In particular, almost sure convergence in concert with the underlying measure space on which the inferential results are considered is crucial for scientific significance beyond just statistical significance.  To be precise, many Machine Learning (ML) and Artificial Intelligence (AI) algorithms and computational packages which can implement them often, though not always, sacrifice interpretability for Model Fit and Prediction. Occam's Razor is perhaps the most well known addage that comes to mind in such contexts, since the more complex we make our model, the harder it is for us to interpret its results. Consider for example, AI and ML methods such as Neural Networks (NN) or Support Vector Machines (SVM). The former may contain multiple hidden layers based on basis expansions of functions that defy any scientific foundation for its existence and the latter suffers from the same predicament depending on the various assumptions used in the model specification. In particular, say we want to predict the heights of certain individuals, and implement a model with sin(log(abs(temperature of a certain lake in antarctica))) such that it perfectly classifies each individual in the Training Data (TrD). Then inspite of such good Model Fit or Prediction (MP) results in the TrD, it is hard to find a reasonable scientific explanation as to why in the general population such a model would predict height well.

Thus, it is reasonable to expect that model variables should be correlated in some scientific manner, which goes beyond just prediction in TrDs, as TeD results and interpretability may be equally important. Therefore, solely relying on one of the many information criteria for TrDs, such as Akaike Information Criteria (AIC) or Bayesian Information Criteria (BIC), does not necessarily guarantee better prediction results for TeDs. This is a fact relevant even if the underlying assumptions of the model specification on the TrD remains true for the TeD, since the presence of unknown latent relationships may become apparent only for the TeD, but not the TrD. Existing learning algorithms, Supervised or Unsupervised, can overcome some of these weaknesses, yet they may suffer from interpretability and overfitting beyond MP results, since the underlying calculations may be Blackboxes without scientifically relevant functional interpretations.

On the other hand, using scientifically interpretable models by itself does not always guarantee the best MP results for TrDs or TeDs either. This is because the existing explanatory variables in the sample may be insufficient to capture the functional specification at all sample points leading to interpretable parameter estimates, yet poor MP results for TeDs. A fact which is also relevant for the methodologies presented here. This is because all such models are a function of the observed Xs. 

Therefore, conventional wisdom generally recognizes that the goal of the analysis should decide the type of methodology that should be used. So if one cares about classification, then we may use one of the many excellent existing AI or ML algorithms separately if interpretability is the goal, we may instead use a simpler functional specification. Yet the connection between MIP for any model considered should provide a more coherent framework for comparison across diffent model and estimation processes considered. Therefore, it would be ideal to go beyond just the Likelihood Principle or information specific criteria, relying on asymptotic results, which themselves are reliant on large samples, to have methodologies that optimize MIP results on all dimensions. 

The issues regarding scientific and statistical significance is of course multifaceted depending invariably on the explanatory variables present, measurement issues, scientfic question of interest, in addition to the mathematical assumptions made in the model specification. Yet an often overlooked criteria is the convergence properties of the methodologies used in model estimations. Since clearly we cannot have an infinite number of sample points, large sample results may be susceptible to both the sample size as well as the estimation procedure (please see \cite{chow21} for further discussion on the parameter bias that may result from this exclusion). Thus, it should not be surprising that violations of the assumptions on which our model is built may give poor MIP results. Unfortunately, Blackbox learning algorithms can do the job for us only partially without prespecified restrictions on the model due to lack of interpretability.

The proposed methodologies through the use of LAHEML, overcomes these existing shortcomings and gives consistent results subject to the model specified and converges to the true parameters under general circumstances. This way the model parameters will converge to the best possible value given the data and the particular model specified. Thus, if the parameters converge to the true values for the model and the model is a priori known, interpretability can be asserted in a rigorous mathematical manner. In so doing, we may also maximize the MP criteria, as a function of the particular model specified and the observed data, without needing to learn it explicitly in some mysterious unobservable way sacrificing interpretability in the process. If the MP results are not deemed to be adequate, the Mathematician can then consider more complex models, perhaps one with interaction terms as opposed to only polynomials of lower orders. As the data change or are updated the model would remain interpretable, being updated to be more complex, but in a known and interpretable way.

As an example, many scientific models are based on convergence in distribution or convergence in probability, on underlying topological spaces which are ``weaker.'' Thus, it suffices to state that conclusions drawn from weaker topological spaces, even if strong convergence is asserted, are only partially informative in comparison to a methodology that asserts almost sure convergence on a stronger topological space. Therefore, the scientific conclusions and inference that we may draw from a methodology that relies on such properties should accordingly be better as well.  

Indeed, this work shows that the conventional wisdom of treating Classification and Inference as separate tasks may not always be necessary. This is because an interpretable model using the right methodology with stronger convergence properties applied to stronger topological spaces can give equivalent or better results than Blackbox AI and ML methods in many circumstances. However, that is not to say that existing methods cannot be used, especially when the scientific question does not require Inference or Prediction at the same time per se. However, if Inference and interpretability are goals, especially when applications of them in the methodologies presented may give similar results to existing AI and ML models, good modeling philosophy should require that they be used first.

Such a philosophy has a long and illustrious history from Aristotle to Occam's Razor in the sciences. Afterall, if we believe in the saying that all models are wrong, but some are useful, must we not then rely on robust methodologies with strong mathematical foundations over reliance on overly complex models? In essence, it highlights that relying on MP criteria to improve Inference and Prediction (IP) may be jointly achievable at the same time under specific mathematical preliminaries, in the constructions presented here. That is, a scientifically interpretable model with robust topological foundations with strong convergence properties can be extremely useful for classification without losing inferential characteristics for the proposed methodologies. On the other hand, a simple model applied without these robust methodologies can lead to the wrong conclusions over existing AI and ML methods, though possibly at the cost of interpretability.

So what then should be the modeling philosophy to go beyond relying on p-values for the sciences? Well unfortunately, there is still much we have to learn. However, we may still be able to say some interesting things given the methodological contributions here. Chief among these is that relying solely on AIC or BIC or other Model Fit criteria might give an incomplete picture, if the data do not conform to the subtle mathematical assumptions of a model. Afterall, there are an infinite number of models one can run on a dataset, therefore, relying solely on minimizing model fit criteria can be just a little time consuming. Relying solely on Prediction criteria can also lead us down the ``Rabbit Hole'' since the performance on TrD and TeD need to be considered carefully. In either case, relying on performance criteria for any one category of modeling objectives does not guarantee that the results will be scientifically interpretable, even if the p-values are small for a predictor or the confidence intervals are amenable to claiming significance.

Therefore, the mathematical results here suggest that a scientifically interpretable model may have excellent predictive capabilities without sacrificing model fit or inference. However, in order to apply such a model, one must consider the convergence properties of the estimation procedure and certain subtle connections between topological spaces and measure spaces. What is important is to note that almost sure estimation methodologies such as LAHEML used in conjunction with stronger topological spaces may give excellent predictive results without sacrificing interpretability of parameter estimates. These results highlight a modeling exercise to be tied to the model specification, and not necessarily entirely dependent on large-sample results or Blackbox learning processes. Therefore, use of LAHEML in scientifically interpretable models can be a first step, which may be sequentially made more complex as necessary irrespective of the statistical goals. Furthermore, if MP is the desired goal, existing AI and ML models may accordingly be improved with these more robust methodologies that have strong convergence properties on stronger topological spaces, since intepretability would no longer be a constraint. In all such cases, methodologies that ensure almost sure convergence is to be preferred over methodologies with other convergence properties. In addition, all such models should be preferred when applied to stronger topological spaces in conjunction with almost sure convergence, to give truly\footnote{Please note that the ideas discussed here are taken from my Dissertation and have been submitted in various forms for publication in various formulations to different sources over the course of my doctoral studies.} ``The Best of Both Worlds!'' 
	
	\section{Conclusion}\label{conclusion}
	
	In summary, this paper presents the most generalized form of the methodologies. It has all the advantages of the existing methodologies and also expands on others. It therefore provides the ideal foundation on which to build any number of supervised or unsupervised methodologies in either the Frequentist or Bayesian formulation. As such, it provides further insights into our continuing discussion on the interplay of scientific significance and statistical significance broadly across scientific fields.
	%

	
	\bibliographystyle{apalike}
	
	\bibliography{References.bib}

\begin{thebibliography}{}

\bibitem[Baldi et~al., 2014]{baldi2014searching}
Baldi, P., Sadowski, P., and Whiteson, D. (2014).
\newblock Searching for exotic particles in high-energy physics with deep
  learning.
\newblock {\em Nature communications}, 5(1):1--9.

\bibitem[Cameron and Trivedi, 2010]{cameron2010microeconometrics}
Cameron, A.~C. and Trivedi, P.~K. (2010).
\newblock {\em Microeconometrics using stata}, volume~2.
\newblock Stata press College Station, TX.

\bibitem[Chowdhury, 2021a]{chow21}
Chowdhury, K. (2021a).
\newblock Functional analysis of generalized linear models under non-linear
  constraints with applications to identifying highly-cited papers.
\newblock {\em Journal of Informetrics}, 15(1):101--112.

\bibitem[Chowdhury, 2021b]{sdss21}
Chowdhury, K. e.~a. (2021b).
\newblock Nonparametric application of functional analysis of generalized
  linear models under nonlinear constraints.
\newblock In {\em Symposium on Data Science and Statistics}. American
  Statistical Association.

\bibitem[Chowdhury, 2021c]{nonparafuncarxiv}
Chowdhury, K.~P. (2021c).
\newblock Nonparametric functional analysis of generalized linear models under
  nonlinear constraints.
\newblock {\em Arxiv}, 2110.04998(04998).

\bibitem[Chowdhury, 2021d]{stronglyconvergentMestarxiv}
Chowdhury, K.~P. (2021d).
\newblock Robust strongly convergent m-estimators under non-iid assumption.
\newblock {\em Arxiv}, 2110.12526(12526).

\bibitem[Chowdhury, 2021e]{mestimators}
Chowdhury, K.~P. (2021e).
\newblock Robust strongly convergent m-estimators under non-iid assumption.
\newblock {\em TBD}.

\bibitem[Draper, 1995]{draper1995assessment}
Draper, D. (1995).
\newblock Assessment and propagation of model uncertainty.
\newblock {\em Journal of the Royal Statistical Society: Series B
  (Methodological)}, 57(1):45--70.

\bibitem[Kass and Steffey, 1989]{kass1989approximate}
Kass, R.~E. and Steffey, D. (1989).
\newblock Approximate bayesian inference in conditionally independent
  hierarchical models (parametric empirical bayes models).
\newblock {\em Journal of the American Statistical Association},
  84(407):717--726.

\bibitem[Killian et~al., 2019]{killian2019learning}
Killian, J.~A., Passino, K.~M., Nandi, A., Madden, D.~R., Clapp, J.~D.,
  Wiratunga, N., Coenen, F., and Sani, S. (2019).
\newblock Learning to detect heavy drinking episodes using smartphone
  accelerometer data.
\newblock In {\em KHD@ IJCAI}, pages 35--42.

\bibitem[Moro et~al., 2014]{moro2014data}
Moro, S., Cortez, P., and Rita, P. (2014).
\newblock A data-driven approach to predict the success of bank telemarketing.
\newblock {\em Decision Support Systems}, 62:22--31.

\bibitem[Tanner and Wong, 1987]{tanner1987calculation}
Tanner, M.~A. and Wong, W.~H. (1987).
\newblock The calculation of posterior distributions by data augmentation.
\newblock {\em Journal of the American statistical Association},
  82(398):528--540.

\end{thebibliography}
\end{document}